\documentclass[twocolumn,aps,amsmath,amssymb,floatfix,prl,superscriptaddress]{revtex4}

\usepackage{bbm}
\usepackage[dvips]{graphicx}
\usepackage{color}
\usepackage{amsmath,amssymb,latexsym}

\newcommand{\Drot}{D_\mathrm{rot}}
\newcommand{\SNR}{\mathrm{SNR}}
\newcommand{\e}{\mathbf{e}}
\newcommand{\n}{\mathbf{n}}
\newcommand{\h}{\mathbf{h}}
\renewcommand{\r}{\mathbf{r}}
\newcommand{\R}{\mathbf{R}}
\newcommand{\s}{\mathbf{s}}
\renewcommand{\t}{\mathbf{t}}
\newcommand{\T}{\mathcal{T}}

\newcommand{\x}{\mathbf{x}}
\newcommand{\stag}{\mathrm{(S)}}
\newcommand{\itag}{\mathrm{(I)}}
\newcommand{\micron}{\mu\mathrm{m}}
\newcommand{\ballistic}{\infty} 
\newcommand{\steady}{\ast} 

\begin{document}

\title{\color{black} Chemokinetic scattering, trapping, and avoidance of active Brownian particles \\
(accepted for publication in Phys. Rev. Lett.)}

\author{Justus A. Kromer}
\affiliation{Stanford University, Stanford, California, United States of America}
\author{Noelia de la Cruz}
\affiliation{McGill University, Montreal, Quebec, Canada}
\author{Benjamin M. Friedrich}
\email{benjamin.m.friedrich@tu-dresden.de}
\affiliation{TU Dresden, Dresden, Germany}

\date{\today}


\begin{abstract} 
We present a theory of chemokinetic search agents that regulate directional fluctuations according to distance from a target. 
A dynamic scattering effect reduces the probability to penetrate regions with high fluctuations
and thus search success for agents that respond instantaneously to positional cues. 
In contrast, agents with internal states that initially suppress chemokinesis can exploit scattering 
to increase their probability to find the target.
Using matched asymptotics between the case of diffusive and ballistic search, we obtain analytic results beyond Fox' colored noise approximation.
\end{abstract}

\maketitle

Many motile cells can navigate in concentration gradients of signaling molecules in a process termed chemotaxis \cite{Fraenkel1940}, 
which guides foraging bacteria to food patches, immune cells to inflammation sites, or sperm cells to the egg. 
Both chemotaxis close to targets 
and random search in the absence of guidance cues 
have each been intensively studied, 
see \cite{Benichou2011,Viswanathan2011,Alvarez2014} for reviews.
Yet, navigation at intermediate distances from a target, 
where chemical cues provide no directional information but only indicate the proximity of a target, 
have attained less attention.
The regulation of speed and persistence of motion as function of absolute concentration of signaling molecules
is known as \textit{chemokinesis} \cite{Fraenkel1940}. 
Chemokinesis 
offers a promising navigation strategy for artificial microrobots with minimal information processing capabilities 
\cite{Mijalkov2016,Nava2018}. 

In biological cells, chemotaxis and chemokinesis usually occur together, making it difficult to disentangle their effects.
At the microscopic scale of cells, molecular shot noise compromises cellular concentration measurements,
rendering cellular steering responses stochastic at low chemoattractant concentrations \cite{Berg1977}. 
We can decompose stochastic steering responses as a superposition of 
directed steering and position-dependent directional fluctuations.

\begin{figure}
\includegraphics[width=1\linewidth]{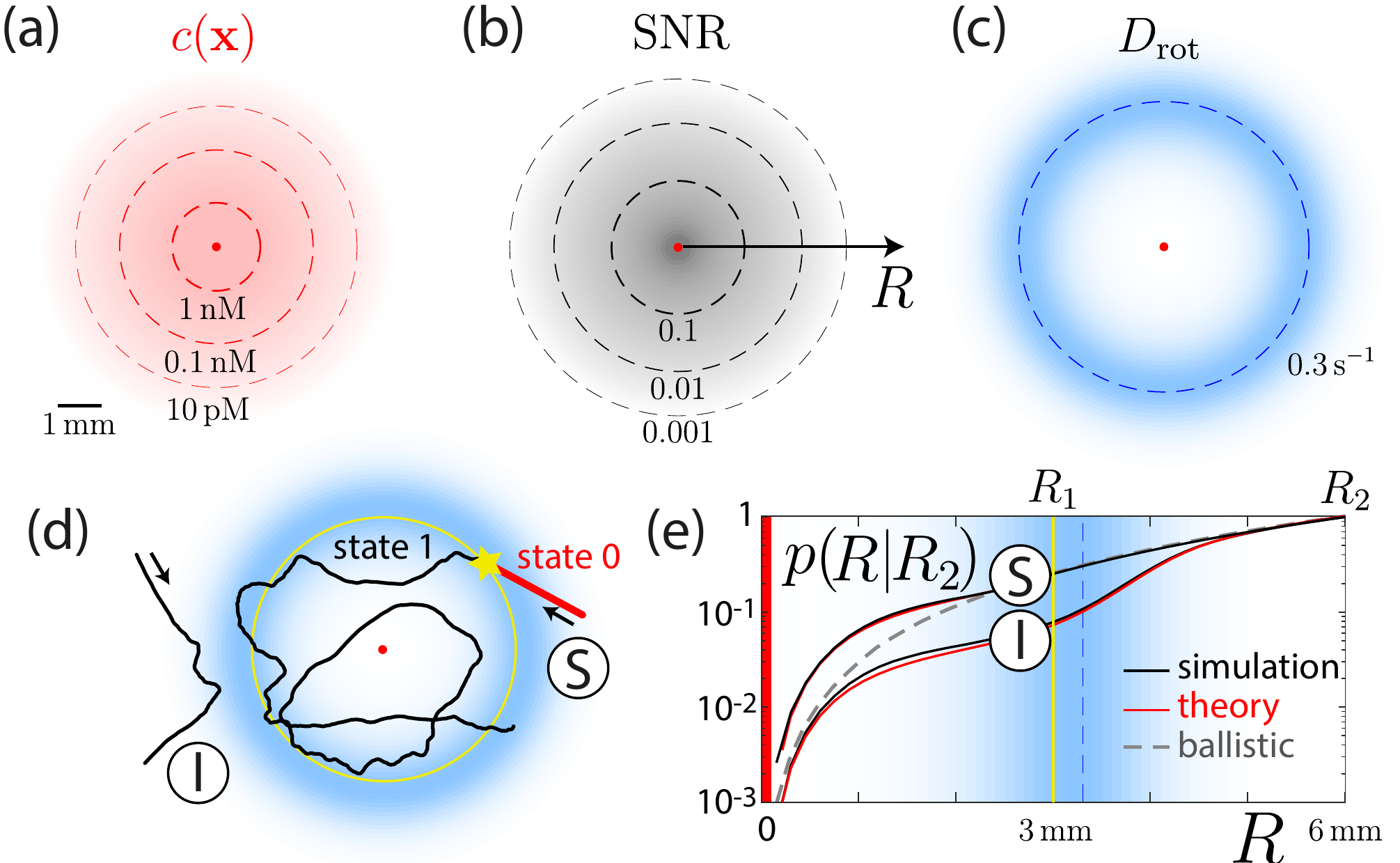}
\caption{
\textbf{Chemokinesis with position-dependent directional fluctuations as consequence of chemotaxis at low concentrations.}
(a)
Radial concentration field $c(\x)$ established by diffusion from a source
(using parameters for sperm chemotaxis \cite{Kromer2018});
source (red dot), iso-concentration lines (dashed).
(b) 
Computed signal-to-noise ratio $\SNR$ of chemotaxis decreases as function of radial distance $R$ from the source. 
(c) 
The corresponding effective rotational diffusion coefficient $\Drot$ displays a maximum at a characteristic distance (dashed),
where $c(\x)$ is still large, but $\SNR$ is low.
(d)
Simulated trajectory [labeled (I)] of a chemokinetic agent subject to $\Drot(\x)$ from panel (c), 
corresponding to \textit{instantaneous chemokinesis}.
A second agent [labeled (S)] with \textit{two-state chemokinesis} initially moves ballistically (state~0), 
but switches to chemokinesis (state~1) as in (I) once it reached a threshold distance for the first time (yellow).
Shown are two-dimensional reconstructions of three-dimensional trajectories 
obtained from numerical integration of Eqs.~(\ref{eq:rPsiDynamics1},\ref{eq:rPsiDynamics2}).
(e)
Penetration probability $p(R|R_2)$ that an ABP starting at distance $R_2$ 
reaches distance $R$ before returning to $R_2$.
For instantaneous chemokinesis (I), this probability is lower than for ballistic motion (dashed).
For two-state chemokinesis (S), $p(R|R_2)$ is higher.
Simulation results (black) compare favorably to our analytic theory (red). 
Parameters, see SM.
}
\label{figure0}
\end{figure}

As illustration, we consider a typical chemotaxis scenario,
sperm cells of marine invertebrates \cite{Jikeli2015}.
There, the egg releases a chemoattractant, which establishes a radial concentration field $c(\x)$ by diffusion \cite{Kromer2018}, see Fig.~\ref{figure0}(a).
Sperm cells can estimate the direction of the local concentration gradient $\nabla c$, 
yet the signal-to-noise ratio $\mathrm{SNR}\sim |\nabla c|^2/c$ of gradient-sensing decreases as function of radial distance $R=|\x|$, 
see Fig.~\ref{figure0}(b).
A previous, generic model of chemotaxis in the presence of sensing noise predicts stochastic steering responses with position-dependent directional fluctuations 
characterized by an effective rotational diffusion coefficient $\Drot(\x)$ \cite{Kromer2018}, see Fig.~\ref{figure0}(c).
Remarkably, $\Drot\sim c/(c+c_b)^2$ becomes maximal at a characteristic distance from the target 
{\color{black}
($\Drot\approx 0.3\,\mathrm{s}^{-1}$ at $R\approx 3.3\,\mathrm{mm}$),
}
marking a `noise zone' 
that incoming cells have to cross \cite{Hein2015,Kromer2018}.
At this distance, absolute chemoattractant concentrations are above the threshold $\color{black} c_b\sim 10\,\mathrm{pM}$ for sensory adaptation, yet $\mathrm{SNR}\ll 1$
[for details, see \cite{Kromer2018} or Supplemental Material (SM)].

Motivated by this example, 
we pose the question whether position-dependent directional fluctuations are beneficial or disadvantageous to find a target.
This question is general:
Spatial modulations of speed or directional fluctuations occur also 
in spatially inhomogeneous activity fields that influence the active motion of artificial microswimmers \cite{Merlitz2017}, 
or from the presence of obstacles \cite{Chepizhko2013,Wang2017}.
Recent studies suggest an intriguing effect of 
position-dependent motility parameters on search success \cite{Schwarz2016, Merlitz2017}, 
termed `pseudochemotaxis' \cite{Schnitzer1993} in \cite{Ghosh2013,Vuijk2018}.

We emphasize that regulation of speed $v=v(\x)$ as function of position $\x$ (termed \textit{orthokinesis} \cite{Fraenkel1940}; considered previously in \cite{Ghosh2013,Vuijk2018}), and %
regulation of effective rotational diffusion coefficient $\Drot=\Drot(\x)$ (\textit{klinokinesis} \cite{Fraenkel1940}; considered here) 
are equivalent:  
We can map orthokinesis on klinokinesis and vice versa, 
by a position-dependent time reparametrization of trajectories proportional to $v(\x)^{-1}$. 
Such reparametrization changes conditional mean first passage times, 
but not the probability to find a target.

In this letter,
we develop a theory of chemokinetic search agents that regulate the level of directional fluctuations as function of distance from a target.
Our model generalizes active Brownian particles (ABP),
frequently used as minimal model for cell motility, 
e.g.\ of biological or artificial microswimmers \cite{Romanczuk2012, BechingerRevModPhys, Friedrich2008}.
We characterize a dynamic scattering effect that reduces the probability to penetrate regions with high fluctuations.
Using matched asymptotics between the limit cases of ballistic and diffusive motion, we develop an analytical theory of this scattering effect.
Scattering always reduces the probability to find a target compared to pure ballistic search for agents that respond instantaneously to positional cues. 
Yet, scattering substantially increases search success for agents with internal states that are able to suppress chemokinesis 
until they came close to the target for a first time,
allowing these agents to realize multiple attempts to hit the target. 
The statistical physics of agents with instantaneous response and those with internal states is fundamentally different: 
while the former display a homogeneous mean residence time, this property is violated in the presence of internal states.

\paragraph{Adaptive Active Brownian Particles (ABP).}

We consider an ABP moving along a trajectory $\textbf{R}(t)$ in three-dimensional space
with speed $v$ and rotational diffusion coefficient $D_\mathrm{rot}$.
Rotational diffusion causes its tangent $\mathbf{t}=\dot{\mathbf{R}}/v$ 
to decorrelate on a time-scale $\tau_p=l_p/v$ set by the persistence length $l_p=v/(2 D_\mathrm{rot})$, 
where dots denote time derivatives.
Hence,
$\langle \mathbf{t}(t_0)\cdot\mathbf{t}(t_0+t) \rangle = \exp ( -|t|/\tau_p)$
\cite{Daniels1952}.
As minimal model of chemokinesis with instantaneous regulation of motility, 
we consider ABP that adjust speed and rotational diffusion coefficient as function of position $\x$, $v=v(\x)$ and $\Drot=\Drot(\mathbf{x})$.
The steady-state density distribution for an ensemble of ABP 
is independent of $\Drot$ and inversely proportional to $v$
(i.e.,\ agents spend proportionally more time in locations, where they move slower), 
with isotropically distributed tangent directions.

Let a single spherical target of radius $R_0$ be located at $\R{=}0$, 
and $v=v(|\x|)$, $\Drot=\Drot(|\x|)$.
Due to spherical symmetry,
the time-dependent distance $R(t)=|\mathbf{R}(t)|$ of the ABP from the origin, 
and the time-dependent angle $\psi(t)$ enclosed by the tangent $\mathbf{t}$ 
and the radial direction $\mathbf{e}_R=-\mathbf{R}/R$, 
decouple from other coordinates, see SM
\begin{eqnarray}
\label{eq:rPsiDynamics1}
\dot{R} &=& - v \cos\psi, \\
\label{eq:rPsiDynamics2}
\dot{\psi} &=& \frac{v}{R}\, \sin\psi +\sqrt{2 \Drot}\, \xi(t) + \Drot\, \mathrm{cot}\,\psi.
\end{eqnarray}   
Here, $\xi(t)$ is Gaussian white noise with
$\langle \xi(t) \rangle=0$ and $\langle \xi(t)\xi(t') \rangle=\delta(t-t')$.

\paragraph{Example: directional fluctuations of chemotaxis.}
We consider an adaptive ABP with constant speed $v$ and position-dependent $\Drot(\x)$ as depicted in Fig.~\ref{figure0}(c).
We assume $R(t=0)=R_2$ and random initial directions with direction angle $\psi$ distributed according $p(\psi)=\sin(2\psi)$ for $0\le\psi\le\pi/2$, 
corresponding to the steady-state influx of ABP at $R_2$ for random initial conditions outside $R_2$, see SM. 

In Fig.~\ref{figure0}(d), the trajectory labeled (I) is scattered back as soon as it encounters an elevated $\Drot$.
Indeed, the penetration probability $p(R|R_2)$ for such ABP starting at distance $R_2$ to reach $R$ before returning to $R_2$ 
is substantially lower than for ballistic motion with $\Drot=0$, see Fig.~\ref{figure0}(e).
For this case of \textit{instantaneous chemokinesis}, 
directional fluctuations reduce the probability $p(R_0|R_2)$ to find the target.

In contrast, we may consider an ABP with two internal states
[labeled (S) in Fig.~\ref{figure0}(d)], 
which initially moves ballistically with $\Drot=0$ (state 0), and only upon crossing a boundary at $R_1$
switches on chemokinesis with $\Drot=\Drot(\x)$ as in case (I) (state 1).
For this \textit{two-state chemokinesis}, 
directional fluctuations increase the probability to find a target, see Fig.~\ref{figure0}(e).
Next, we consider minimal models to explain this phenomenon.

\paragraph{Spatial inhomogeneous directional fluctuations cause dynamic scattering of ABP.}

\begin{figure}
\includegraphics[width=1\linewidth]{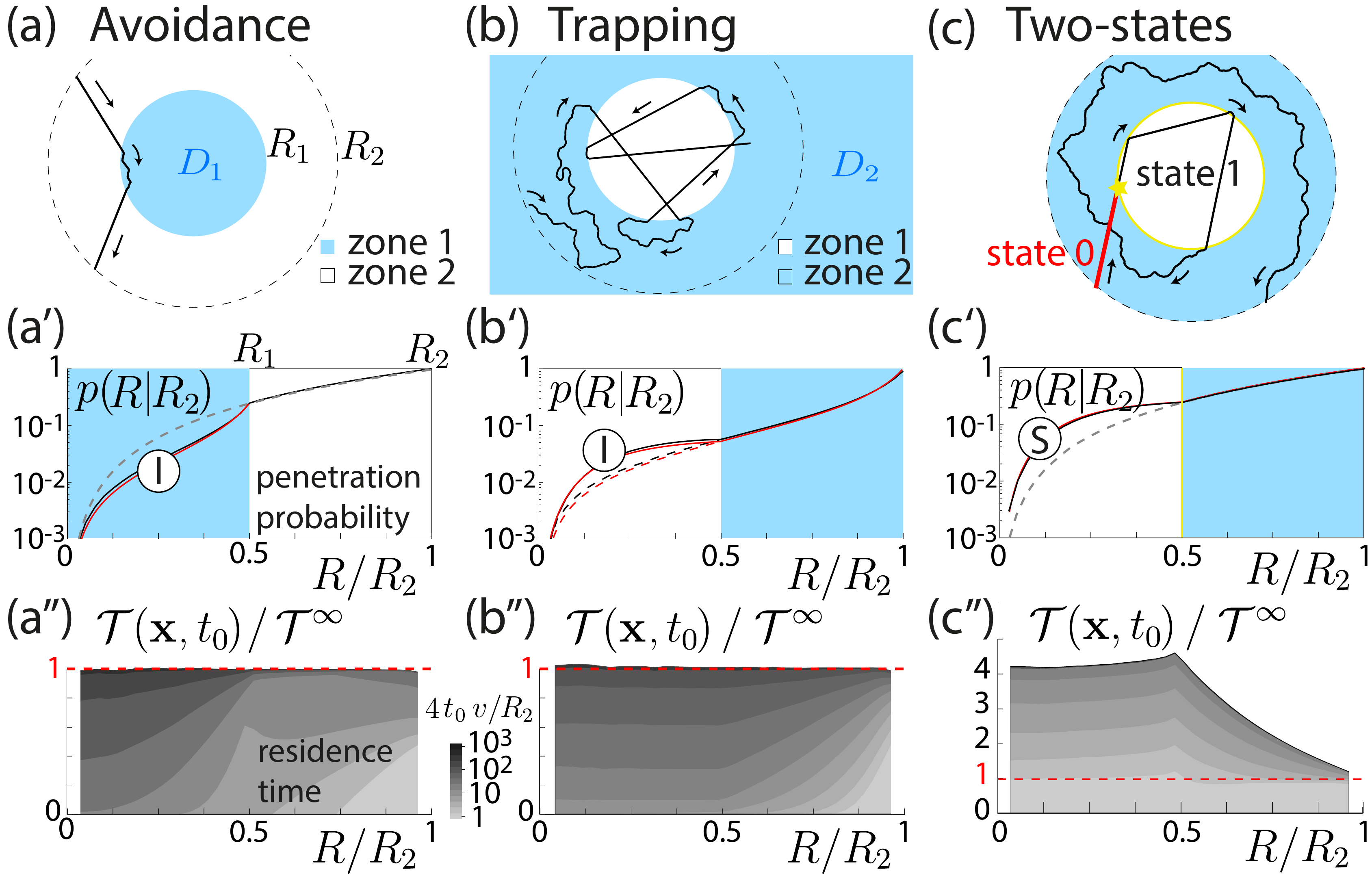}
\caption{
\textbf{Dynamic scattering of chemokinetic agents.}
(a)
Trajectories entering zone 1 (blue) are scattered back to zone 2 due to 
a high rotational diffusion coefficient $D_1{>}0$ in zone 1. 
(a$'$): 
Penetration probability $p(R|R_2)$
that an ABP starting at distance $R_2$ reaches distance $R$ before returning to $R_2$, 
analogous to Fig.~\ref{figure0}(e): simulation (black), theory (red), ballistic motion with $D_1=D_2=0$ (dashed).
(a$''$) 
Residence time $\T(\x,t_0)$ at space position $\x$ before time $t_0$ as function of radial distance $R{=}|\x|$.
By a mean-chord-length theorem, 
$\lim_{t_0\rightarrow\infty} \T(\x,t_0) = \T^\ballistic=(\pi R_2^2 v)^{-1}$.
(b) Same as panel (a), but with $D_1{=}0$, $D_2{>}0$:
trajectories can become trapped in zone 1 due to inward scattering of outgoing trajectories at $R_1$.
(c) Same as panel (b), but for two-state chemokinesis:
agents initially move ballistically with $\Drot{=}0$ (state~0), 
and switch to chemokinesis with $\Drot=\Drot(\x)$ as in (b) after crossing a threshold distance (yellow) at $R_1$ (state~1).
(c$'$) The penetration probability is higher compared to ballistic motion.
(c$''$) The mean residence time is now spatially inhomogeneous.
Parameters: $R_1{=}R_2/2$; 
(a): $D_1{=}10 v/R_2$, $D_2{=}0$,
(b): $D_1{=}0$, $D_2{=}10 v/R_2$; reference case (dashed) in (b$'$): $D_1{=}D_2{=}10 v/R_2$.
}
\label{figure1}
\end{figure}

We first consider a minimal model with constant speed $v$ 
and rotational diffusion coefficient $D_\mathrm{rot}(R)$
that is piecewise constant in zones concentric with the target, see Fig.~\ref{figure1}(a)
\begin{equation}
\label{eq:PersistenceLength}
D_\mathrm{rot}(R)= 
\begin{cases}
D_1    & \text{for } R<R_1    \text{ (zone 1)}\\
D_2    & \text{for } R\ge R_1 \text{ (zone 2)}
\end{cases}.
\end{equation}

We illustrate the effect of spatially inhomogeneous rotational diffusion coefficient
in two special cases, termed \textit{avoidance} and \textit{trapping} \cite{Fraenkel1940}, see Fig.~\ref{figure1}(a,b).
ABPs start at $R{=}R_2$ 
with random inward pointing initial direction angles $\psi$ 
and terminate once they reach $R_2$ again. 

If the ABP increases $\Drot$ upon entering zone 1, 
most trajectories that enter zone 1 promptly return to zone 2,
being scattered back due to the decrease in directional persistence, see Fig.~\ref{figure1}(a).

Again, the penetration probability $p(R|R_2)$ for this case is lower than for ballistic motion, see Fig.~\ref{figure1}(a$'$):
most ABP avoid zone 1.
However, the ensemble-averaged residence time $\T(\x)$ at each position $\x$ (with units time-per-volume) is spatially homogeneous,
and equals the value $\T^\ballistic$ for ballistic motion.
This is a direct corollary of the fact that the steady-state probability density for Eqs.~(\ref{eq:rPsiDynamics1},\ref{eq:rPsiDynamics2}) is independent of $\Drot$.
Elementary geometry gives $\T^\ballistic = 4/(v S)$ with $S{=}4\pi R_2^2$ \cite{Case1967}.

Thus, 
the mean residence time of \textit{inhomogeneous} persistent random walks 
is the same as the mean residence time for ballistic motion.
This extends a prominent result for \textit{homogeneous} stochastic motion \cite{Blanco2003,Benichou2005b},
also known as mean-chord-length property,
which found applications for wave scattering \cite{Pierrat2014} and modeling of neutron transport \cite{Zoia2019}.
The original proof can be adapted to inhomogeneous stochastic motion, asserted in \cite{Blanco2006}.
Related results were discussed for position-dependent translational diffusion \cite{Schnitzer1993,Lau2007,Nava2018}.

Intuitively, although most trajectories are reflected away from zone 1, 
a small fraction of trajectories will penetrate into zone 1 and dwell there an extended period of time before leaving eventually.
Fig.~\ref{figure1}(a$''$) shows a time-bounded residence time $\T(\x,t_0)$ 
to find an ABP at position $\x$ at distance $R$ before time $t_0$
[with $\lim_{t_0\rightarrow\infty} \T(\x,t_0)=\T(\x)$].

If the ABP instead increases $\Drot$ when leaving zone 1, 
trajectories that have just left zone 1 may be scattered back, see Fig.~\ref{figure1}(b).
ABPs are ``trapped'' in zone 1.
Concomitantly, $p(R|R_2)$ is higher 
than for spatially homogeneous persistent random walks with $D_1=D_2>0$, see Fig.~\ref{figure1}(b$'$).
Again, $\T(\x)=\T^\ballistic$, see Fig.~\ref{figure1}(b$''$). 
Intuitively, although some trajectories become trapped, many trajectories are scattered back to $R_2$ before they ever enter zone 1. 

The case in Fig.~\ref{figure1}(a) corresponds to chemokinetic avoidance \cite{Fraenkel1940};
imagine, zone 1 represents unfavorable conditions that agents seek to avoid. 
Remarkably, while most agents benefit, 
a small number will suffer an adverse effect, spending more time in the unfavorable zone~1.

\paragraph{Instantaneous chemokinesis $\mathrm{(I)}$ versus two-state chemokinesis $\mathrm{(S)}$.}

\begin{figure}
\includegraphics[width=\linewidth]{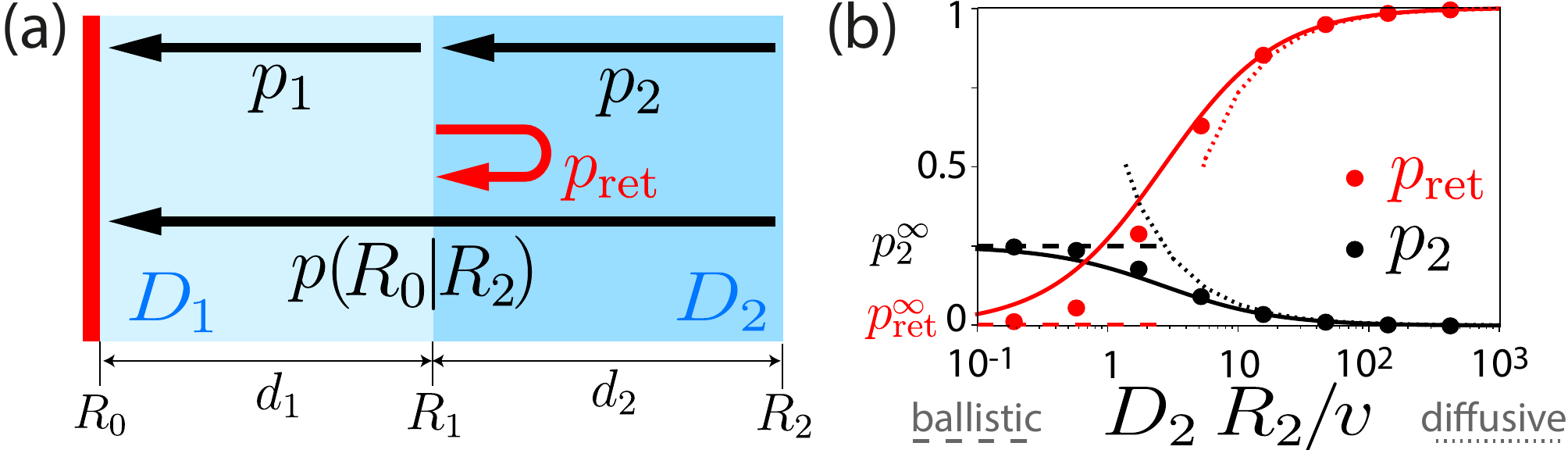}
\caption{
\textbf{Analytic theory by matched asymptotics.}
(a): The probabilities $p_1$ and $p_2$ to pass zone $1$ and $2$, respectively, 
and the return probability $p_\mathrm{ret}$ at the boundary between zone 1 and 2, 
jointly determine the probability $p(R_0|R_2)$ to reach a target of radius $R_0$
for ABP starting at $R_2$, see Eq.~(\ref{eq:pin}).
(b): Trade-off between $p_2$ (black) and $p_\mathrm{ret}$ (red) as function of $D_2$.
Simulation (dots); 
limit of high $\Drot$ (dotted);
ballistic motion [Eq.~(\ref{eq:past})] (dashed);
matched asymptotics [Eq.~(\ref{eq:prefl})] (solid).
}
\label{figure2}
\end{figure}

To characterize the role of scattering for target search, 
we introduce the return probability $p_\mathrm{ret}$ to re-enter zone 1 
after entering zone 2 at $R{=}R_1$ (with random outwards pointing initial direction), 
and analogous zone-crossing probabilities $p_1{=}p(R_0|R_1)$ and $p_2{=}p(R_1|R_2)$,
see Fig.~\ref{figure2}(a).
The probability $p(R_0|R_2)$ for ABP with instantaneous chemokinesis starting at $R_2$ to hit the target of radius $R_0$
can be expressed in terms of these zone-crossing probabilities as a geometric series
\begin{equation}
\label{eq:pin}
p(R_0|R_2) \approx \sum_{k=0}^\infty p_2 [(1-p_1)p_\mathrm{ret}]^k p_1.
\end{equation}
Here, the $k$-th summand denotes the probability
of successful trajectories that cross $R_1$ exactly $2k+1$-times.
The only assumption made in deriving Eq.~(\ref{eq:pin}) is a stereotypic distribution of direction angles at zone boundaries.
Eq.~(\ref{eq:pin}) corroborates that inward scattering at $R_1$
implies effective trapping of trajectories in zone 1, 
allowing for multiple attempts to hit the target.

Generally, $p_i$ is a monotonically decreasing function of $D_i$, see Fig.~\ref{figure2}(b), 
with maximal value $p_i^\ballistic$ obtained for ballistic motion
\begin{equation}
\label{eq:past}
p_i^\ballistic = \lim_{D_i \rightarrow 0} p_i
         = \left( \frac{R_{i-1}}{R_i} \right)^2, \quad i=1,2. 
\end{equation}
Note that $p_\mathrm{ret}$ and $p_2$ both depend on $D_2$, 
and thus cannot be optimized independently, see Fig.~\ref{figure2}(b):
increasing $D_2$ increases scattering of outgoing trajectories (thus increasing $p_\mathrm{ret}$),
yet also increases scattering of incoming trajectories (thus decreasing $p_2$).

An ABP with two internal states can decouple 
scattering of incoming and outgoing trajectories.
Analogous to Fig.~\ref{figure0}(d)-label~(S),
we assume that ABP initially move ballistically with $\Drot{=}0$ (state~0).
Upon first entering zone 1, ABP permanently switch to state~1 
and subsequently obey Eq.~(\ref{eq:PersistenceLength}), see Fig.~\ref{figure1}(c).
Fig.~\ref{figure1}(c$'$) demonstrates a dramatic increase of $p(R|R_2)$.
Concomitantly, $\T(\x)$ is not homogeneous anymore, see Fig.~\ref{figure1}(c$''$).

\paragraph{Analytical theory.}

We derive approximate analytical expressions for the return probability $p_\mathrm{ret}$ 
using matched asymptotics ($p_1$ and $p_2$ are analogous).
Results compare favorably to simulations, see Fig.~\ref{figure1}.

An ABP entering zone 2 from zone 1 at time $t{=}0$
initially continues moving in approximately radial direction,
before its direction of motion decorrelates on a time-scale $\tau_p{=}(2D_2)^{-1}$.
For times $t\gg\tau_p$, the ABP exhibits isotropic random motion.
We treat these two dynamic phases separately, 
and introduce a cross-over time $t_0$ with $\tau_p\ll t_0\ll D_2 (d_2/v_0)^2$.

For the first phase,
we are interested in the \textit{penetration depth}
$\langle x(t)\rangle$,
i.e.,\ the conditional expectation value 
of radial position $R_1+x$
of ABPs that have not yet been absorbed at $R_1$ at time $t$.
Let $Q(t)$ be the corresponding survival probability.
In the limit $l_2\ll R_1$, 
we can approximate the absorbing spherical shell at $R_1$ by a plane $H$.
Using symmetry of renewal processes under reflection at $H$, 
we compute 
$\lim_{t\rightarrow\infty} \langle x(t)\rangle Q(t) = \alpha l_2$ with $\alpha=4/3$, see SM.
Intuitively, while fewer and fewer ABPs survive,
their mean distance from $H$ diverges as $\langle x(t)\rangle \sim \alpha l_2 (t/\tau_p)^{1/2}$.

We now address the second dynamic phase $t\ge t_0$, 
and calculate $p_\mathrm{ret}$.
Those ABPs that have not been absorbed at $R_1$ before $t_0$
will likely be found at a distance $x\gg l_2$ from $R_1$, 
and we may approximate these as diffusive particles.
The probability that a diffusive particle reaches $R_2$ if released at radial position $R$
between two absorbing spherical shells of radii $R_1$ and $R_2$ reads 
$p(R)=(R_2/R)(R-R_1)/(R_2-R_1)$ 
\cite{Berg1977}.
Choosing
$R=R_1+\langle x(t_0)\rangle \approx R_1+\alpha l_2/Q(t_0)$
yields an asymptotic result for 
$q=1-p_\mathrm{ret}$
as
$q \approx 
Q(t_0)p[R(t_0)] 
$, 
valid for 
$\lambda_2\ll (\tau_p/t_0)^{1/2}$.
Here, we introduced the ratios $\lambda_i=l_i/d_i$ between the persistence length $l_i=v/(2D_i)$ inside zone $i$ and 
zone width $d_i=R_i-R_{i-1}$, $i=1,2$. 

We can extend this asymptotic expression to the entire range $0<\lambda_2<\infty$ by interpolating 
with the limit value 
$q^\ballistic=\lim_{\lambda_2\rightarrow\infty}1-p_\mathrm{ret}=1$
for ballistic motion, 
using the simple ansatz of a saturation curve 
$q \approx 
\gamma\lambda_2 q^\ballistic / [ q^\ballistic+\gamma\lambda_2 ]$ with 
$\gamma = {\partial q / \partial\lambda_2}_{|\lambda_2=0}$.
We find
\begin{equation}
\label{eq:prefl}
p_\mathrm{ret} \approx 
\left( 1 + \alpha \lambda_2\,R_2/R_1 \right)^{-1}, \quad \alpha=4/3.
\end{equation}
Analogously, 
$p_i \approx \alpha\lambda_i\,p_i^\ballistic R_{i-1} / (R_i p_i^\ballistic+\alpha\lambda_iR_{i-1})$
with $\lambda_i=v/(2 D_i d_i)$, $i=1,2$.

\paragraph{Continuum limit.}
By induction, we can generalize
the minimal model of Eq.~(\ref{eq:PersistenceLength}) with $n=2$ zones
to the case of $n>2$ zones concentric with the origin bounded by $R_{i-1}<R\le R_i$, $i=1\ldots,n$. 
From Eqs.~(\ref{eq:pin},\ref{eq:prefl}), we obtain a recursion relation for $p(R_0|R_i)$, see SM for details.
In the continuum limit $n\rightarrow\infty$, 
we obtain a differential equation for $p(R_0|R)$, 
describing the penetration probability for the case of instantaneous chemokinesis,
where $\Drot=\Drot(R)$ may be an arbitrary function of $R$
\begin{equation}
\frac{\partial}{\partial R} p(R_0|R) = - 2 \frac{p}{R} - \Drot(R)\,\frac{3}{2}\frac{p^2}{v} .
\end{equation}
By definition, $p(R_0|R_0)=1$.
We conclude that for instantaneous chemokinesis,
the probability to find a target is always smaller compared to ballistic motion. 
For two-state chemokinesis with threshold distance at $R_1$, 
the penetration probability 
equals $p^\ballistic(R|R_2)$ for $R_1\le R<R_2$, 
and differs from $p(R|R_2)$ by a factor $p^\ballistic(R_1|R_2) / p(R_1|R_2)>1$ for $R<R_1$, 
see Fig.~\ref{figure1}(c$'$).

\paragraph{Discussion.}

Using a minimal model of a chemokinetic agent 
that regulates its rotational diffusion coefficient $\Drot$ as function of distance from a target, 
we explain how search agents can harness spatially inhomogeneous directional fluctuations 
to find targets more efficiently if they possess internal states.

This chemokinesis strategy exploits 
a dynamic scattering effect 
that scatters agents away from regions where directional fluctuations are high.
Agents that have just missed the target and move on an outgoing trajectory
can thus become scattered inward again and realize an additional search attempt.
Yet, agents with instantaneous chemokinesis face a trade-off between 
this beneficial inward scattering of outgoing trajectories, and 
unwanted outward scattering of incoming trajectories.
Agents can avoid this trade-off if they suppress chemokinesis when approaching the target for the first time. 
In our minimal model, this strategy is realized by agents with two internal states,
{\color{black}
which could be implemented by a bistable switch, see SM;
}
alternatively, sensorial delay, memory, or hysteresis could serve a similar purpose.

Scattering is a genuinely dynamic effect.
Consequently, 
the probability to find a target in spatially inhomogeneous systems cannot be predicted 
from mass-action laws on the basis of ensemble-averaged mean residence times
(which in fact are spatially homogeneous for instantaneous chemokinesis).
This highlights a fundamental difference between the dynamical and the steady-state behavior 
of spatially inhomogeneous active systems \cite{Vuijk2018,Ghosh2013}. 

The dynamic scattering effect described here explains the increased target encounter rates previously observed 
for spatially-heterogeneous search of particles switching between ballistic and diffusive runs \cite{Schwarz2016,Schwarz2016b}, 
as well as search in spatially inhomogeneous activity fields \cite{Merlitz2017,Vuijk2018}
(using the mapping between klinokinesis and orthokinesis, see introduction).


Our work connects to a recent interest in composite search strategies
\cite{Plank2008,Loverdo2009,Benichou2011,Bartumeus2014,Nolting2015}. 
While most authors considered agents that stochastically switch between different levels of directional fluctuations,
switching is triggered by proximity to a target in our case, 
representing \textit{resource-sensitive} composite search \cite{Benhamou1992,Nolting2015}.

In addition to chemokinesis studied here, 
chemotaxis can become useful in the ultimate vicinity of the target, 
where the signal-to-noise ratio of gradient-sensing exceeds one, 
thus setting an effective target size.
%
{\color{black}
Our theoretical work suggests that 
single-molecule sensitivity of chemotactic cells \cite{Strunker2015} may in fact be disadvantageous
during the initial approach to a target surrounded by a static, radial concentration field.
In contrast, single-molecule sensitivity would be advantageous  
after cells have passed a `noise zone',
where the concentration of signaling molecules equals the cell's sensitivity threshold. 
As experimental test,
chemotactic responses of cells with single-molecule sensitivity
could be compared before and after exposure to high concentrations.
}

\begin{acknowledgments}
JAK and BMF are supported by the German National Science Foundation (DFG) 
through the Excellence Initiative by the German Federal and State Governments 
(Clusters of Excellence cfaed EXC-1056 and PoL EXC-2068),
as well as DFG grant FR3429/3-1 to BMF.
NdC acknowledges a RISE-Globalink Research Internship.
We thank Rainer Klages, Jens-Uwe Sommer, and Steffen Lange for a critical reading of the manuscript.
\end{acknowledgments}

\bibliography{chemokinesis}

\begin{thebibliography}{40}
\expandafter\ifx\csname natexlab\endcsname\relax\def\natexlab#1{#1}\fi
\expandafter\ifx\csname bibnamefont\endcsname\relax
  \def\bibnamefont#1{#1}\fi
\expandafter\ifx\csname bibfnamefont\endcsname\relax
  \def\bibfnamefont#1{#1}\fi
\expandafter\ifx\csname citenamefont\endcsname\relax
  \def\citenamefont#1{#1}\fi
\expandafter\ifx\csname url\endcsname\relax
  \def\url#1{\texttt{#1}}\fi
\expandafter\ifx\csname urlprefix\endcsname\relax\def\urlprefix{URL }\fi
\providecommand{\bibinfo}[2]{#2}
\providecommand{\eprint}[2][]{\url{#2}}

\bibitem[{\citenamefont{Fraenkel and Gunn}(1961)}]{Fraenkel1940}
\bibinfo{author}{\bibfnamefont{G.}~\bibnamefont{Fraenkel}} \bibnamefont{and}
  \bibinfo{author}{\bibfnamefont{D.}~\bibnamefont{Gunn}},
  \emph{\bibinfo{title}{The Orientation of Animals}} (\bibinfo{publisher}{Dover
  Publ., New York}, \bibinfo{year}{1961}).

\bibitem[{\citenamefont{B{\'e}nichou et~al.}(2011)\citenamefont{B{\'e}nichou,
  Loverdo, Moreau, and Voituriez}}]{Benichou2011}
\bibinfo{author}{\bibfnamefont{O.}~\bibnamefont{B{\'e}nichou}},
  \bibinfo{author}{\bibfnamefont{C.}~\bibnamefont{Loverdo}},
  \bibinfo{author}{\bibfnamefont{M.}~\bibnamefont{Moreau}}, \bibnamefont{and}
  \bibinfo{author}{\bibfnamefont{R.}~\bibnamefont{Voituriez}},
  \bibinfo{journal}{Rev. Mod. Phys.} \textbf{\bibinfo{volume}{83}},
  \bibinfo{pages}{81} (\bibinfo{year}{2011}).

\bibitem[{\citenamefont{Viswanathan et~al.}(2011)\citenamefont{Viswanathan,
  Da~Luz, Raposo, and Stanley}}]{Viswanathan2011}
\bibinfo{author}{\bibfnamefont{G.~M.} \bibnamefont{Viswanathan}},
  \bibinfo{author}{\bibfnamefont{M.~G.~E.} \bibnamefont{Da~Luz}},
  \bibinfo{author}{\bibfnamefont{E.~P.} \bibnamefont{Raposo}},
  \bibnamefont{and} \bibinfo{author}{\bibfnamefont{H.~E.}
  \bibnamefont{Stanley}}, \emph{\bibinfo{title}{The {P}hysics of {F}oraging}}
  (\bibinfo{publisher}{Cambridge Univ. Press}, \bibinfo{year}{2011}).

\bibitem[{\citenamefont{Alvarez et~al.}(2014)\citenamefont{Alvarez, Friedrich,
  Gompper, and Kaupp}}]{Alvarez2014}
\bibinfo{author}{\bibfnamefont{L.}~\bibnamefont{Alvarez}},
  \bibinfo{author}{\bibfnamefont{B.~M.} \bibnamefont{Friedrich}},
  \bibinfo{author}{\bibfnamefont{G.}~\bibnamefont{Gompper}}, \bibnamefont{and}
  \bibinfo{author}{\bibfnamefont{U.~B.} \bibnamefont{Kaupp}},
  \bibinfo{journal}{Trends Cell Biol.} \textbf{\bibinfo{volume}{24}},
  \bibinfo{pages}{198} (\bibinfo{year}{2014}).

\bibitem[{\citenamefont{Mijalkov et~al.}(2016)\citenamefont{Mijalkov, McDaniel,
  Wehr, and Volpe}}]{Mijalkov2016}
\bibinfo{author}{\bibfnamefont{M.}~\bibnamefont{Mijalkov}},
  \bibinfo{author}{\bibfnamefont{A.}~\bibnamefont{McDaniel}},
  \bibinfo{author}{\bibfnamefont{J.}~\bibnamefont{Wehr}}, \bibnamefont{and}
  \bibinfo{author}{\bibfnamefont{G.}~\bibnamefont{Volpe}},
  \bibinfo{journal}{Phys. Rev. X} \textbf{\bibinfo{volume}{6}},
  \bibinfo{pages}{011008} (\bibinfo{year}{2016}).

\bibitem[{\citenamefont{Nava et~al.}(2018)\citenamefont{Nava, Gro{\ss}mann, and
  Peruani}}]{Nava2018}
\bibinfo{author}{\bibfnamefont{L.~G.} \bibnamefont{Nava}},
  \bibinfo{author}{\bibfnamefont{R.}~\bibnamefont{Gro{\ss}mann}},
  \bibnamefont{and} \bibinfo{author}{\bibfnamefont{F.}~\bibnamefont{Peruani}},
  \bibinfo{journal}{Phys. Rev. E} \textbf{\bibinfo{volume}{97}},
  \bibinfo{pages}{042604} (\bibinfo{year}{2018}).

\bibitem[{\citenamefont{Berg and Purcell}(1977)}]{Berg1977}
\bibinfo{author}{\bibfnamefont{H.~C.} \bibnamefont{Berg}} \bibnamefont{and}
  \bibinfo{author}{\bibfnamefont{E.~M.} \bibnamefont{Purcell}},
  \bibinfo{journal}{Biophys. J.} \textbf{\bibinfo{volume}{20}},
  \bibinfo{pages}{193} (\bibinfo{year}{1977}).

\bibitem[{\citenamefont{Jikeli et~al.}(2015)\citenamefont{Jikeli, Alvarez,
  Friedrich, Wilson, Pascal, Colin, Pichlo, Rennhack, Brenker, and
  Kaupp}}]{Jikeli2015}
\bibinfo{author}{\bibfnamefont{J.~F.} \bibnamefont{Jikeli}},
  \bibinfo{author}{\bibfnamefont{L.}~\bibnamefont{Alvarez}},
  \bibinfo{author}{\bibfnamefont{B.~M.} \bibnamefont{Friedrich}},
  \bibinfo{author}{\bibfnamefont{L.~G.} \bibnamefont{Wilson}},
  \bibinfo{author}{\bibfnamefont{R.}~\bibnamefont{Pascal}},
  \bibinfo{author}{\bibfnamefont{R.}~\bibnamefont{Colin}},
  \bibinfo{author}{\bibfnamefont{M.}~\bibnamefont{Pichlo}},
  \bibinfo{author}{\bibfnamefont{A.}~\bibnamefont{Rennhack}},
  \bibinfo{author}{\bibfnamefont{C.}~\bibnamefont{Brenker}}, \bibnamefont{and}
  \bibinfo{author}{\bibfnamefont{U.~B.} \bibnamefont{Kaupp}},
  \bibinfo{journal}{Nat. Commun.} \textbf{\bibinfo{volume}{6}},
  \bibinfo{pages}{7985} (\bibinfo{year}{2015}).

\bibitem[{\citenamefont{Kromer et~al.}(2018)\citenamefont{Kromer,
  M{\"{a}}rcker, Lange, Baier, and Friedrich}}]{Kromer2018}
\bibinfo{author}{\bibfnamefont{J.}~\bibnamefont{Kromer}},
  \bibinfo{author}{\bibfnamefont{S.}~\bibnamefont{M{\"{a}}rcker}},
  \bibinfo{author}{\bibfnamefont{S.}~\bibnamefont{Lange}},
  \bibinfo{author}{\bibfnamefont{C.}~\bibnamefont{Baier}}, \bibnamefont{and}
  \bibinfo{author}{\bibfnamefont{B.~M.} \bibnamefont{Friedrich}},
  \bibinfo{journal}{PLoS Comp. Biol.} \textbf{\bibinfo{volume}{14}},
  \bibinfo{pages}{1} (\bibinfo{year}{2018}).

\bibitem[{\citenamefont{Hein et~al.}(2016)\citenamefont{Hein, Brumley, Carrara,
  Stocker, and Levin}}]{Hein2015}
\bibinfo{author}{\bibfnamefont{A.~M.} \bibnamefont{Hein}},
  \bibinfo{author}{\bibfnamefont{D.~R.} \bibnamefont{Brumley}},
  \bibinfo{author}{\bibfnamefont{F.}~\bibnamefont{Carrara}},
  \bibinfo{author}{\bibfnamefont{R.}~\bibnamefont{Stocker}}, \bibnamefont{and}
  \bibinfo{author}{\bibfnamefont{S.~A.} \bibnamefont{Levin}},
  \bibinfo{journal}{J. Roy. Soc. Interf.} \textbf{\bibinfo{volume}{13}},
  \bibinfo{pages}{20150844} (\bibinfo{year}{2016}).

\bibitem[{\citenamefont{Merlitz et~al.}(2017)\citenamefont{Merlitz, Wu, and
  Sommer}}]{Merlitz2017}
\bibinfo{author}{\bibfnamefont{H.}~\bibnamefont{Merlitz}},
  \bibinfo{author}{\bibfnamefont{C.}~\bibnamefont{Wu}}, \bibnamefont{and}
  \bibinfo{author}{\bibfnamefont{J.-U.} \bibnamefont{Sommer}},
  \bibinfo{journal}{Soft Matter} \textbf{\bibinfo{volume}{13}},
  \bibinfo{pages}{3726} (\bibinfo{year}{2017}).

\bibitem[{\citenamefont{Chepizhko and Peruani}(2013)}]{Chepizhko2013}
\bibinfo{author}{\bibfnamefont{O.}~\bibnamefont{Chepizhko}} \bibnamefont{and}
  \bibinfo{author}{\bibfnamefont{F.}~\bibnamefont{Peruani}},
  \bibinfo{journal}{Phys. Rev. Lett.} \textbf{\bibinfo{volume}{111}},
  \bibinfo{pages}{160604} (\bibinfo{year}{2013}).

\bibitem[{\citenamefont{Wang et~al.}(2017)\citenamefont{Wang, Zhang, Xia, and
  Yu}}]{Wang2017}
\bibinfo{author}{\bibfnamefont{J.}~\bibnamefont{Wang}},
  \bibinfo{author}{\bibfnamefont{D.}~\bibnamefont{Zhang}},
  \bibinfo{author}{\bibfnamefont{B.}~\bibnamefont{Xia}}, \bibnamefont{and}
  \bibinfo{author}{\bibfnamefont{W.}~\bibnamefont{Yu}}, \bibinfo{journal}{Soft
  Matter} \textbf{\bibinfo{volume}{13}}, \bibinfo{pages}{758}
  (\bibinfo{year}{2017}).

\bibitem[{\citenamefont{Schwarz
  et~al.}(2016{\natexlab{a}})\citenamefont{Schwarz, Schr{\"o}der, Qu, Hoth, and
  Rieger}}]{Schwarz2016}
\bibinfo{author}{\bibfnamefont{K.}~\bibnamefont{Schwarz}},
  \bibinfo{author}{\bibfnamefont{Y.}~\bibnamefont{Schr{\"o}der}},
  \bibinfo{author}{\bibfnamefont{B.}~\bibnamefont{Qu}},
  \bibinfo{author}{\bibfnamefont{M.}~\bibnamefont{Hoth}}, \bibnamefont{and}
  \bibinfo{author}{\bibfnamefont{H.}~\bibnamefont{Rieger}},
  \bibinfo{journal}{Phys. Rev. Lett.} \textbf{\bibinfo{volume}{117}},
  \bibinfo{pages}{068101} (\bibinfo{year}{2016}{\natexlab{a}}).

\bibitem[{\citenamefont{Schnitzer}(1993)}]{Schnitzer1993}
\bibinfo{author}{\bibfnamefont{M.~J.} \bibnamefont{Schnitzer}},
  \bibinfo{journal}{Phys. Rev. E} \textbf{\bibinfo{volume}{48}},
  \bibinfo{pages}{2553} (\bibinfo{year}{1993}).

\bibitem[{\citenamefont{Ghosh et~al.}(2013)\citenamefont{Ghosh, Misko,
  Marchesoni, and Nori}}]{Ghosh2013}
\bibinfo{author}{\bibfnamefont{P.~K.} \bibnamefont{Ghosh}},
  \bibinfo{author}{\bibfnamefont{V.~R.} \bibnamefont{Misko}},
  \bibinfo{author}{\bibfnamefont{F.}~\bibnamefont{Marchesoni}},
  \bibnamefont{and} \bibinfo{author}{\bibfnamefont{F.}~\bibnamefont{Nori}},
  \bibinfo{journal}{Phys. Rev. Lett.} \textbf{\bibinfo{volume}{110}},
  \bibinfo{pages}{268301} (\bibinfo{year}{2013}).

\bibitem[{\citenamefont{Vuijk et~al.}(2018)\citenamefont{Vuijk, Sharma, Mondal,
  Sommer, and Merlitz}}]{Vuijk2018}
\bibinfo{author}{\bibfnamefont{H.~D.} \bibnamefont{Vuijk}},
  \bibinfo{author}{\bibfnamefont{A.}~\bibnamefont{Sharma}},
  \bibinfo{author}{\bibfnamefont{D.}~\bibnamefont{Mondal}},
  \bibinfo{author}{\bibfnamefont{J.-U.} \bibnamefont{Sommer}},
  \bibnamefont{and} \bibinfo{author}{\bibfnamefont{H.}~\bibnamefont{Merlitz}},
  \bibinfo{journal}{Phys. Rev. E} \textbf{\bibinfo{volume}{97}},
  \bibinfo{pages}{042612} (\bibinfo{year}{2018}).

\bibitem[{\citenamefont{Romanczuk et~al.}(2012)\citenamefont{Romanczuk,
  B{\"a}r, Ebeling, Lindner, and Schimansky-Geier}}]{Romanczuk2012}
\bibinfo{author}{\bibfnamefont{P.}~\bibnamefont{Romanczuk}},
  \bibinfo{author}{\bibfnamefont{M.}~\bibnamefont{B{\"a}r}},
  \bibinfo{author}{\bibfnamefont{W.}~\bibnamefont{Ebeling}},
  \bibinfo{author}{\bibfnamefont{B.}~\bibnamefont{Lindner}}, \bibnamefont{and}
  \bibinfo{author}{\bibfnamefont{L.}~\bibnamefont{Schimansky-Geier}},
  \bibinfo{journal}{Eur. Phys. J. Spec. Top.} \textbf{\bibinfo{volume}{202}},
  \bibinfo{pages}{1} (\bibinfo{year}{2012}).

\bibitem[{\citenamefont{Bechinger et~al.}(2016)\citenamefont{Bechinger,
  Di~Leonardo, L\"owen, Reichhardt, Volpe, and Volpe}}]{BechingerRevModPhys}
\bibinfo{author}{\bibfnamefont{C.}~\bibnamefont{Bechinger}},
  \bibinfo{author}{\bibfnamefont{R.}~\bibnamefont{Di~Leonardo}},
  \bibinfo{author}{\bibfnamefont{H.}~\bibnamefont{L\"owen}},
  \bibinfo{author}{\bibfnamefont{C.}~\bibnamefont{Reichhardt}},
  \bibinfo{author}{\bibfnamefont{G.}~\bibnamefont{Volpe}}, \bibnamefont{and}
  \bibinfo{author}{\bibfnamefont{G.}~\bibnamefont{Volpe}},
  \bibinfo{journal}{Rev. Mod. Phys.} \textbf{\bibinfo{volume}{88}},
  \bibinfo{pages}{045006} (\bibinfo{year}{2016}).

\bibitem[{\citenamefont{Friedrich}(2008)}]{Friedrich2008}
\bibinfo{author}{\bibfnamefont{B.~M.} \bibnamefont{Friedrich}},
  \bibinfo{journal}{Phys. Biol.} \textbf{\bibinfo{volume}{5}},
  \bibinfo{pages}{026007} (\bibinfo{year}{2008}).

\bibitem[{\citenamefont{Daniels}(1952)}]{Daniels1952}
\bibinfo{author}{\bibfnamefont{H.~E.} \bibnamefont{Daniels}},
  \bibinfo{journal}{Proc. Roy. Soc. Edinb. A} \textbf{\bibinfo{volume}{63}},
  \bibinfo{pages}{290–311} (\bibinfo{year}{1952}).

\bibitem[{\citenamefont{Case and Zweifel}(1967)}]{Case1967}
\bibinfo{author}{\bibfnamefont{K.~M.} \bibnamefont{Case}} \bibnamefont{and}
  \bibinfo{author}{\bibfnamefont{P.~F.} \bibnamefont{Zweifel}},
  \emph{\bibinfo{title}{Linear Transport Theory}}
  (\bibinfo{publisher}{Addison-Wesley, Reading}, \bibinfo{year}{1967}).

\bibitem[{\citenamefont{Blanco and Fournier}(2003)}]{Blanco2003}
\bibinfo{author}{\bibfnamefont{S.}~\bibnamefont{Blanco}} \bibnamefont{and}
  \bibinfo{author}{\bibfnamefont{R.}~\bibnamefont{Fournier}},
  \bibinfo{journal}{EPL} \textbf{\bibinfo{volume}{61}}, \bibinfo{pages}{168}
  (\bibinfo{year}{2003}).

\bibitem[{\citenamefont{B{\'e}nichou et~al.}(2005)\citenamefont{B{\'e}nichou,
  Coppey, Moreau, Suet, and Voituriez}}]{Benichou2005b}
\bibinfo{author}{\bibfnamefont{O.}~\bibnamefont{B{\'e}nichou}},
  \bibinfo{author}{\bibfnamefont{M.}~\bibnamefont{Coppey}},
  \bibinfo{author}{\bibfnamefont{M.}~\bibnamefont{Moreau}},
  \bibinfo{author}{\bibfnamefont{P.}~\bibnamefont{Suet}}, \bibnamefont{and}
  \bibinfo{author}{\bibfnamefont{R.}~\bibnamefont{Voituriez}},
  \bibinfo{journal}{EPL} \textbf{\bibinfo{volume}{70}}, \bibinfo{pages}{42}
  (\bibinfo{year}{2005}).

\bibitem[{\citenamefont{Pierrat et~al.}(2014)\citenamefont{Pierrat, Ambichl,
  Gigan, Haber, Carminati, and Rotter}}]{Pierrat2014}
\bibinfo{author}{\bibfnamefont{R.}~\bibnamefont{Pierrat}},
  \bibinfo{author}{\bibfnamefont{P.}~\bibnamefont{Ambichl}},
  \bibinfo{author}{\bibfnamefont{S.}~\bibnamefont{Gigan}},
  \bibinfo{author}{\bibfnamefont{A.}~\bibnamefont{Haber}},
  \bibinfo{author}{\bibfnamefont{R.}~\bibnamefont{Carminati}},
  \bibnamefont{and} \bibinfo{author}{\bibfnamefont{S.}~\bibnamefont{Rotter}},
  \bibinfo{journal}{Proc. Natl. Acad. Sci. U.S.A.}
  \textbf{\bibinfo{volume}{111}}, \bibinfo{pages}{17765}
  (\bibinfo{year}{2014}).

\bibitem[{\citenamefont{Zoia et~al.}(2019)\citenamefont{Zoia, Larmier, and
  Mancusi}}]{Zoia2019}
\bibinfo{author}{\bibfnamefont{A.}~\bibnamefont{Zoia}},
  \bibinfo{author}{\bibfnamefont{C.}~\bibnamefont{Larmier}}, \bibnamefont{and}
  \bibinfo{author}{\bibfnamefont{D.}~\bibnamefont{Mancusi}},
  \bibinfo{journal}{EPL} \textbf{\bibinfo{volume}{127}}, \bibinfo{pages}{20006}
  (\bibinfo{year}{2019}).

\bibitem[{\citenamefont{Blanco and Fournier}(2006)}]{Blanco2006}
\bibinfo{author}{\bibfnamefont{S.}~\bibnamefont{Blanco}} \bibnamefont{and}
  \bibinfo{author}{\bibfnamefont{R.}~\bibnamefont{Fournier}},
  \bibinfo{journal}{Phys. Rev. Lett.} \textbf{\bibinfo{volume}{97}},
  \bibinfo{pages}{230604} (\bibinfo{year}{2006}).

\bibitem[{\citenamefont{Lau and Lubensky}(2007)}]{Lau2007}
\bibinfo{author}{\bibfnamefont{A.~W.~C.} \bibnamefont{Lau}} \bibnamefont{and}
  \bibinfo{author}{\bibfnamefont{T.~C.} \bibnamefont{Lubensky}},
  \bibinfo{journal}{Phys. Rev. E} \textbf{\bibinfo{volume}{76}},
  \bibinfo{pages}{011123} (\bibinfo{year}{2007}).

\bibitem[{\citenamefont{Schwarz
  et~al.}(2016{\natexlab{b}})\citenamefont{Schwarz, Schr{\"o}der, and
  Rieger}}]{Schwarz2016b}
\bibinfo{author}{\bibfnamefont{K.}~\bibnamefont{Schwarz}},
  \bibinfo{author}{\bibfnamefont{Y.}~\bibnamefont{Schr{\"o}der}},
  \bibnamefont{and} \bibinfo{author}{\bibfnamefont{H.}~\bibnamefont{Rieger}},
  \bibinfo{journal}{Phys. Rev. E} \textbf{\bibinfo{volume}{94}},
  \bibinfo{pages}{042133} (\bibinfo{year}{2016}{\natexlab{b}}).

\bibitem[{\citenamefont{Plank and James}(2008)}]{Plank2008}
\bibinfo{author}{\bibfnamefont{M.~J.} \bibnamefont{Plank}} \bibnamefont{and}
  \bibinfo{author}{\bibfnamefont{A.}~\bibnamefont{James}}, \bibinfo{journal}{J.
  Roy. Soc. Int.} \textbf{\bibinfo{volume}{5}}, \bibinfo{pages}{1077}
  (\bibinfo{year}{2008}).

\bibitem[{\citenamefont{Loverdo et~al.}(2009)\citenamefont{Loverdo, B\'enichou,
  Moreau, and Voituriez}}]{Loverdo2009}
\bibinfo{author}{\bibfnamefont{C.}~\bibnamefont{Loverdo}},
  \bibinfo{author}{\bibfnamefont{O.}~\bibnamefont{B\'enichou}},
  \bibinfo{author}{\bibfnamefont{M.}~\bibnamefont{Moreau}}, \bibnamefont{and}
  \bibinfo{author}{\bibfnamefont{R.}~\bibnamefont{Voituriez}},
  \bibinfo{journal}{Phys. Rev. E} \textbf{\bibinfo{volume}{80}},
  \bibinfo{pages}{031146} (\bibinfo{year}{2009}).

\bibitem[{\citenamefont{Bartumeus et~al.}(2014)\citenamefont{Bartumeus, Raposo,
  Viswanathan, and da~Luz}}]{Bartumeus2014}
\bibinfo{author}{\bibfnamefont{F.}~\bibnamefont{Bartumeus}},
  \bibinfo{author}{\bibfnamefont{E.~P.} \bibnamefont{Raposo}},
  \bibinfo{author}{\bibfnamefont{G.~M.} \bibnamefont{Viswanathan}},
  \bibnamefont{and} \bibinfo{author}{\bibfnamefont{M.~G.}
  \bibnamefont{da~Luz}}, \bibinfo{journal}{PLoS One}
  \textbf{\bibinfo{volume}{9}}, \bibinfo{pages}{e106373}
  (\bibinfo{year}{2014}).

\bibitem[{\citenamefont{Nolting et~al.}(2015)\citenamefont{Nolting, Hinkelman,
  Brassil, and Tenhumberg}}]{Nolting2015}
\bibinfo{author}{\bibfnamefont{B.~C.} \bibnamefont{Nolting}},
  \bibinfo{author}{\bibfnamefont{T.~M.} \bibnamefont{Hinkelman}},
  \bibinfo{author}{\bibfnamefont{C.~E.} \bibnamefont{Brassil}},
  \bibnamefont{and}
  \bibinfo{author}{\bibfnamefont{B.}~\bibnamefont{Tenhumberg}},
  \bibinfo{journal}{Ecol. Complex.} \textbf{\bibinfo{volume}{22}},
  \bibinfo{pages}{126} (\bibinfo{year}{2015}).

\bibitem[{\citenamefont{Benhamou}(1992)}]{Benhamou1992}
\bibinfo{author}{\bibfnamefont{S.}~\bibnamefont{Benhamou}},
  \bibinfo{journal}{J. Theoret. Biol.} \textbf{\bibinfo{volume}{159}},
  \bibinfo{pages}{67} (\bibinfo{year}{1992}).

\bibitem[{\citenamefont{Str{\"u}nker et~al.}(2015)\citenamefont{Str{\"u}nker,
  Alvarez, and Kaupp}}]{Strunker2015}
\bibinfo{author}{\bibfnamefont{T.}~\bibnamefont{Str{\"u}nker}},
  \bibinfo{author}{\bibfnamefont{L.}~\bibnamefont{Alvarez}}, \bibnamefont{and}
  \bibinfo{author}{\bibfnamefont{U.}~\bibnamefont{Kaupp}},
  \bibinfo{journal}{Curr. Opinion Neurobiol.} \textbf{\bibinfo{volume}{34}},
  \bibinfo{pages}{110} (\bibinfo{year}{2015}).

\bibitem[{\citenamefont{Friedrich and J\"ulicher}(2009)}]{Friedrich2009}
\bibinfo{author}{\bibfnamefont{B.~M.} \bibnamefont{Friedrich}}
  \bibnamefont{and}
  \bibinfo{author}{\bibfnamefont{F.}~\bibnamefont{J\"ulicher}},
  \bibinfo{journal}{Phys. Rev. Lett.} \textbf{\bibinfo{volume}{103}},
  \bibinfo{pages}{068102} (\bibinfo{year}{2009}).

\bibitem[{\citenamefont{Sharma et~al.}(2017)\citenamefont{Sharma, Wittmann, and
  Brader}}]{Sharma2017}
\bibinfo{author}{\bibfnamefont{A.}~\bibnamefont{Sharma}},
  \bibinfo{author}{\bibfnamefont{R.}~\bibnamefont{Wittmann}}, \bibnamefont{and}
  \bibinfo{author}{\bibfnamefont{J.~M.} \bibnamefont{Brader}},
  \bibinfo{journal}{Phys. Rev. E} \textbf{\bibinfo{volume}{95}},
  \bibinfo{pages}{012115} (\bibinfo{year}{2017}).

\bibitem[{\citenamefont{Fox}(1986)}]{Fox1986}
\bibinfo{author}{\bibfnamefont{R.}~\bibnamefont{Fox}}, \bibinfo{journal}{Phys.
  Rev. A} \textbf{\bibinfo{volume}{33}}, \bibinfo{pages}{467}
  (\bibinfo{year}{1986}).

\bibitem[{\citenamefont{Kashikar et~al.}(2012)\citenamefont{Kashikar, Alvarez,
  Seifert, Gregor, J{\"a}ckle, Beyermann, Krause, and Kaupp}}]{Kashikar2012}
\bibinfo{author}{\bibfnamefont{N.~D.} \bibnamefont{Kashikar}},
  \bibinfo{author}{\bibfnamefont{L.}~\bibnamefont{Alvarez}},
  \bibinfo{author}{\bibfnamefont{R.}~\bibnamefont{Seifert}},
  \bibinfo{author}{\bibfnamefont{I.}~\bibnamefont{Gregor}},
  \bibinfo{author}{\bibfnamefont{O.}~\bibnamefont{J{\"a}ckle}},
  \bibinfo{author}{\bibfnamefont{M.}~\bibnamefont{Beyermann}},
  \bibinfo{author}{\bibfnamefont{E.}~\bibnamefont{Krause}}, \bibnamefont{and}
  \bibinfo{author}{\bibfnamefont{U.~B.} \bibnamefont{Kaupp}},
  \bibinfo{journal}{J. Cell Biol.} \textbf{\bibinfo{volume}{198}},
  \bibinfo{pages}{1075} (\bibinfo{year}{2012}).

\bibitem[{\citenamefont{Pichlo et~al.}(2014)\citenamefont{Pichlo,
  Bungert-Pluemke, Weyand, Seifert, Boenigk, Str\"{u}nker, Kashikar, Goodwin,
  M\"{u}ller, K\"{o}rschen et~al.}}]{Pichlo2014}
\bibinfo{author}{\bibfnamefont{M.}~\bibnamefont{Pichlo}},
  \bibinfo{author}{\bibfnamefont{S.}~\bibnamefont{Bungert-Pluemke}},
  \bibinfo{author}{\bibfnamefont{I.}~\bibnamefont{Weyand}},
  \bibinfo{author}{\bibfnamefont{R.}~\bibnamefont{Seifert}},
  \bibinfo{author}{\bibfnamefont{W.}~\bibnamefont{Boenigk}},
  \bibinfo{author}{\bibfnamefont{T.}~\bibnamefont{Str\"{u}nker}},
  \bibinfo{author}{\bibfnamefont{N.~D.} \bibnamefont{Kashikar}},
  \bibinfo{author}{\bibfnamefont{N.}~\bibnamefont{Goodwin}},
  \bibinfo{author}{\bibfnamefont{A.}~\bibnamefont{M\"{u}ller}},
  \bibinfo{author}{\bibfnamefont{H.~G.} \bibnamefont{K\"{o}rschen}},
  \bibnamefont{et~al.}, \bibinfo{journal}{J. Cell Biol.}
  \textbf{\bibinfo{volume}{206}}, \bibinfo{pages}{541} (\bibinfo{year}{2014}).

\end{thebibliography}


\clearpage

\appendix

\section{Supplemental Material}

{\noindent
Justus A. Kromer,
Noelia de la Cruz,
Benjamin M. Friedrich:
\textbf{
Chemokinetic scattering, trapping, and avoidance of active Brownian particles
}}

\renewcommand{\theequation}{S\arabic{equation}}    
\setcounter{equation}{0}  
\renewcommand{\thefigure}{S\arabic{figure}}    
\setcounter{figure}{0}  
\renewcommand{\thetable}{S\arabic{table}}    
\setcounter{table}{0}  
\renewcommand{\thepage}{S\arabic{page}}    
\setcounter{page}{1}  

\subsection{Numerical methods}

For numeric integration of Eqs.~(1) and (2), 
we used an explicit Euler-Maruyama method with integration time step $\Delta t = 0.5\cdot 10^{-4}\,R_2/v$.
As control, we additionally simulated ABP trajectories in three-dimensional space, 
using an Euler-Heun scheme with matrix exponentials for propagation of the Frenet-Serret frame.
Time step was $dt=0.25\cdot 10^{-3}\,R_2/v$; 
analogous simulations using a smaller time step of $dt=0.25\cdot 10^{-4}\,R_2/v$ gave consistent results.
ABP were initially positioned at $R_2$ with initial direction angle $\psi$ distributed according to 
$p(\psi)=\sin(2 \psi)$ for $0\le\psi<\pi/2$ and $p(\psi)=0$ else (unless stated otherwise).
Simulations were stopped after a maximum search time of 
$250\,R_2/v$ for Fig.~1(e) and Fig.~2, 
$500\,R_2/v$ for Fig.~3(b), 
$2.4\cdot 10^3\,R_2/v$ for Fig.~\ref{som_fig_vuijk}(a), and
$1.5\cdot 10^5\,D_0^{-1}$ for Fig.~\ref{som_fig_vuijk}(b).
We simulated $n=10^6$ ABP per data point 
for Fig.~2, Fig.~3(b), Fig.~\ref{fig_p_psi}, and Fig.~\ref{som_fig_vuijk}(a), as well as
$n=10^4$ ABP for Fig. 1(e), and Fig.~\ref{som_fig_vuijk}(b).
For Fig.~\ref{som_fig_vuijk}(b) [orthokinesis], an adaptive time-step was used.





\subsection{Derivation of Eqs.~(1,2)}
\label{sec:rpsi_dyn}

The stochastic differential equation Eqs.~(1,2) 
can be derived using previously published ideas \cite{Friedrich2009,Kromer2018}.
We include the derivation for completeness, 
generalizing the derivation in the PhD thesis of one of the authors
(available at: http://nbn-resolving.de/urn:nbn:de:bsz:14-ds-1235056439247-79608).

We consider the trajectory $\R(t)$ of an ABP with speed $v$ and rotational diffusion coefficient $\Drot$,
together with a material frame with orthonormal vectors $\h_1$, $\h_2$, $\h_3$, where $\h_3$ shall denote the tangent $\h_3=\t=\dot{\R}/v$ of $\R(t)$. 
The stochastic equation of motion reads in Stratonovich (S) interpretation
\begin{equation}
\label{eq:motion}
\stag
\begin{array}{rlll}
\dot{\R}    &= v\,\h_3 , & & \\
\quad \dot{\h}_3 &= -           \xi_1\,\h_2 & + \xi_2\,\h_1, & \\
\dot{\h}_1       &=                         & - \xi_2\,\h_3 & + \xi_3\,\h_2 , \\
\dot{\h}_2       &= \phantom{-} \xi_1\,\h_3 &               & - \xi_3\,\h_1 ,
\end{array}
\end{equation}
where $\xi_i(t)$ denote independent white noise processes
with $\langle \xi_i(t)\xi_j(t')\rangle = 2\Drot \,\delta_{ij}\,\delta(t-t')$.

\begin{figure}
\includegraphics[width=\linewidth]{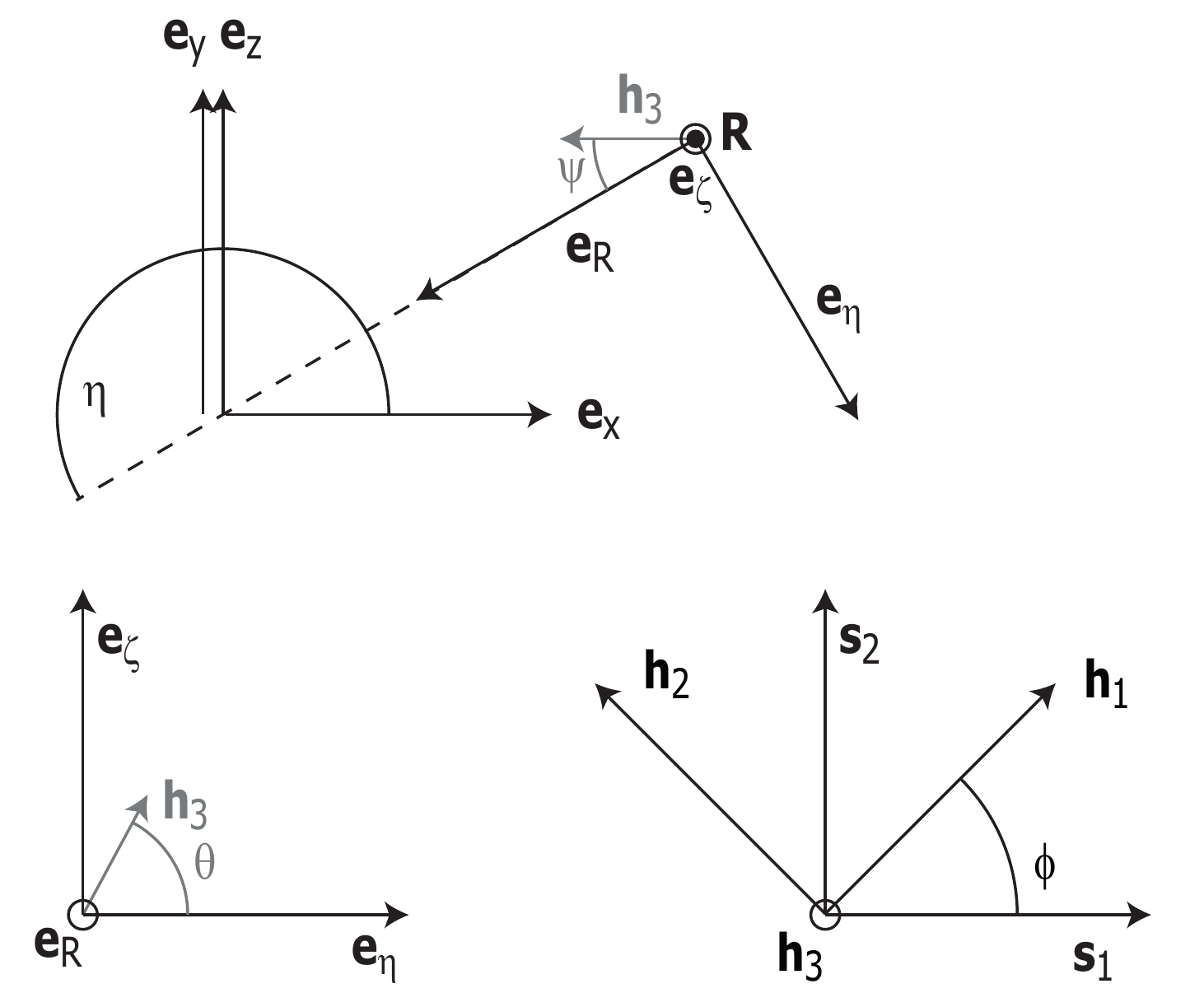}
\caption{
\textbf{Euler angles.} 
Sketch of vectors and Euler angles used in the derivation of Eq.~(2), see text.
}
\label{figure_sketch_spherical_coordinates}
\end{figure}

We will rewrite the equation of motion Eq.~(\ref{eq:motion}) using spherical coordinates, 
see Fig.~\ref{figure_sketch_spherical_coordinates}.
We first introduce a system of orthonormal vectors 
comprising the inward radial vector $\e_R = -\R/R$ with $R=|\R|$, 
as well as vectors $\e_\eta$ and $\e_\zeta$ given by 
\begin{equation}
\begin{split}
\e_R     &= (\phantom{-}\cos\eta,\sin\eta\cos\zeta,\sin\eta\sin\zeta) , \\
\e_\eta  &= (-\sin\eta,\cos\eta\cos\zeta,\cos\eta\sin\zeta) , \\
\e_\zeta &= (0,-\sin\zeta,\cos\zeta) . \\
\end{split}
\end{equation}
We note 
$\e_\eta=\frac{\partial}{\partial\eta} \e_R$, 
and 
$\e_\zeta=\e_R\times\e_\eta=\frac{\partial}{\partial\zeta} \e_R / \sin\eta$. 

We express the position vector $\R$ and the material frame vectors $\h_1$, $\h_2$, $\h_3$ 
with respect to $\e_R$, $\e_\eta$, $\e_\zeta$,
introducing Euler angles $\psi$, $\theta$, $\varphi$
\begin{equation}
\label{eq:spherical_coords}
\begin{array}{rlll}
\R &= -R \,\e_R , & & \\
\h_3 &= \phantom{-} \cos\psi\,\e_R + \sin\psi \left( \cos\theta\,\e_\eta + \sin\theta\,\e_\zeta \right), \\
\h_1 &= \phantom{-} \cos\varphi\, \s_1 + \sin\varphi\, \s_2 , \\
\h_2 &= -           \sin\varphi\, \s_1 + \cos\varphi\, \s_2 = \h_3\times\h_1 , \\
\end{array}
\end{equation}
where
\begin{align}
\s_1 
&= \frac{\partial}{\partial\psi} \h_3 \\
&= -\sin\psi\,\e_R + \cos\psi\, (\cos\theta\,\e_\eta+\sin\theta\,\e_\zeta) , \\
\s_2 &= \h_3\times\s_1 = \frac{\partial}{\partial\theta}\h_3/\sin\psi \\ 
&= -\sin\theta\,\e_\eta + \cos\theta\,\e_\zeta .
\end{align}
Note that $\s_1$, $\s_2$, $\h_3$ form a system of orthonormal vectors;
rotation by $\varphi$ around $\h_3$ maps this system onto $\h_1$, $\h_2$, $\h_3$.

The vectors $\e_R$, $\e_\eta$, $\e_\zeta$ obey the dynamic equations
\begin{equation}
\stag
\begin{array}{rlll}
\dot{\e}_R     &= \phantom{-} \dot{\eta}\, \e_\eta & + \dot{\zeta}\,\sin\eta\,\e_\zeta , & \\
\dot{\e}_\eta  &= -           \dot{\eta}\, \e_R    &                                     & + \dot{\zeta}\,\cos\eta\,\e_\zeta , \\
\dot{\e}_\zeta &=                                  & - \dot{\zeta}\,\sin\eta\,\e_R       & - \dot{\zeta}\,\cos\eta\,\e_\eta , \\
\end{array}
\end{equation}
whereas
\begin{equation}
\stag
\begin{array}{rl}
\dot{\s}_1 &= -\dot{\psi}\,\h_3 + \dot{\theta}\cos\psi\,\s_2 \, \\
& \quad -\sin\psi\,\dot{\e}_R+\cos\psi\,(\cos\theta\,\dot{\e}_\eta+\sin\theta\,\dot{\e}_\zeta) , \\
\dot{\s}_2 &= -\dot{\theta}\,(\sin\psi\,\h_3 + \cos\psi\,\s_1 ) \, \\
&\quad - \sin\theta\,\dot{\e}_\eta + \cos\theta\,\dot{\e}_\zeta .
\end{array}
\end{equation}
Note that the rules of ordinary calculus apply in Stratonovich calculus.

From Eq.~(\ref{eq:spherical_coords}), we find for the following scalar products
\begin{equation}
\label{eq:scalar_products1a}
\stag
\begin{array}{rl}
\dot{\R}\cdot\e_R &=
- \dot{R}, \\
\dot{\R}\cdot\e_\eta &=
-R\,\dot{\eta}, \\
\dot{\R}\cdot\e_\zeta &=
-R\,\dot{\zeta}\,\sin\eta,
\end{array}
\end{equation}
as well as
\begin{equation}
\label{eq:scalar_products1b}
\stag
\begin{array}{rl}
\dot{\h}_3\cdot\e_R &=
-\sin\psi\, (\dot{\psi}+\dot{\eta}\,\cos\theta+\dot{\zeta}\,\sin\theta\sin\eta ), \\
\dot{\h}_3\cdot\e_\eta &=
\cos\psi\,(\dot{\eta}+\dot{\psi}\,\cos\theta)-\sin\psi\sin\theta\,(\dot{\theta}-\dot{\zeta}\,\cos\eta), \\ 
\dot{\h}_1\cdot\h_2 
&= \dot{\varphi} + \dot{\s}_1\cdot\s_2 \\
&= \dot{\varphi} + \dot{\theta}\,\cos\psi + \dot{\eta}\,\sin\psi\sin\theta \, \\
& \quad - \dot{\zeta}\,(\sin\psi\sin\eta\cos\theta - \cos\psi\cos\eta) .
\end{array}
\end{equation}
Using Eq.~(\ref{eq:motion}), we can express these scalar products alternatively as
\begin{equation}
\label{eq:scalar_products2a}
\stag
\begin{array}{rl}
\dot{\R}\cdot\e_R &= 
v \cos\psi , \\
\dot{\R}\cdot\e_\eta &=
v \sin\psi \cos\theta , \\
\dot{\R}\cdot\e_\zeta &=
v \sin\psi\sin\theta , 
\end{array}
\end{equation}
as well as
\begin{equation}
\label{eq:scalar_products2b}
\stag
\begin{array}{rl}
\dot{\h}_3\cdot\e_R &= -\Xi_1\,\sin\psi , \\
\dot{\h}_3\cdot\e_\eta 
&= \phantom{-} \Xi_1\,\cos\psi \cos\theta +\Xi_2\,\sin\theta , \\
\dot{\h}_1\cdot\h_2 &= \xi_3 ,
\end{array}
\end{equation}
where we used short-hand
\begin{align}
\Xi_1 &= \xi_1\,\sin\varphi + \xi_2\,\cos\varphi , \\
\Xi_2 &= \xi_1\,\cos\varphi - \xi_2\,\sin\varphi . 
\end{align}

Comparing Eqs.~(\ref{eq:scalar_products1a},\ref{eq:scalar_products1b}) and (\ref{eq:scalar_products2a},\ref{eq:scalar_products2b}),
we can now solve for $\dot{R}$, $\dot{\eta}$, $\dot{\zeta}$, $\dot{\psi}$, $\dot{\theta}$, $\dot{\varphi}$ and find
\begin{equation}
\stag
\begin{array}{rl}
\dot{R} = & -v\,\cos\psi , \\ 
\dot{\eta} = & -\frac{v}{R} \sin\psi\cos\theta , \\
\dot{\zeta} = & -\frac{v}{R} \sin\psi\sin\theta/\sin\eta , \\
\dot{\psi} = & \phantom{-} \frac{v}{R}\sin\psi + \Xi_1 , \\
\dot{\theta} = & 
- \frac{v}{R} \sin\psi\cos\theta - \Xi_2/\sin\psi, \\
\dot{\varphi} = & \xi_3 + \Xi_2\,\cot\psi \, \\
&\quad +\frac{v}{R}\sin\psi\,(\cos\theta\cos\psi - \sin\theta\cos\psi\cot\eta) .
\end{array}
\end{equation}

The stochastic differential equation for $\dot{\psi}$ contains a multiplicative noise term $\Xi_1$, 
which depends on $\varphi$.
We can decouple the dynamics of $\psi$ from $\varphi$ by using a simple trick: 
First, we rewrite the equation as an It{\=o} stochastic differential equation
\begin{align}
\itag\quad
\dot{\psi} & = \frac{v}{R}\sin\psi + \xi_1\,\sin\varphi + \xi_2\,\cos\varphi + \Drot \cot\psi .
\end{align}
This It\=o differential equation contains a noise-induced drift term \cite{Lau2007},
$(1/2)\,2\Drot\,d(\sin\varphi)/d\varphi \,\cos\varphi \, \cot\psi + (1/2)\,2\Drot\,d(\cos\varphi)/d\varphi \,(-\sin\varphi) \, \cot\psi = \Drot \cot\psi$.
In It{\=o} calculus, $\xi_1\,\sin\varphi+\xi_2\,\cos\varphi$ is equivalent
to a Gaussian white noise term $\xi$ with $\langle \xi(t)\,\xi(t')\rangle = 2\Drot\, \delta(t-t')$
since $\varphi(t)$ and $\xi_i(t)$, $i=1,2$ are independent.
Thus, we have an equivalent Langevin equation for the dynamics of $\psi$, 
which contains only non-multiplicative noise $\xi(t)$
\begin{equation}
\boxed{
\dot{\psi} = \frac{v}{R}\sin\psi + \xi+ \Drot \cot\psi .
}
\end{equation}

The corresponding Fokker-Planck equation reads
\begin{align}
\dot{P}(R,\psi) = 
& -\frac{\partial}{\partial R} \left[ - v\cos\psi\, P \right ] \notag \\
& -\frac{\partial}{\partial \psi} \left[ \left( \frac{v}{R}\sin\psi + \Drot\,\cot\psi\right) \,P \right ] \notag \\
& + \frac{\partial^2}{\partial^2 \psi} \Drot\, P(R,\psi). 
\end{align}
The steady-state density distribution $P^\steady(R,\psi)$ for an ensemble of agents follows as
\begin{equation}
\boxed{
P^\steady(R,\psi) \sim 4\pi\,R^2\,\frac{\sin\psi}{2}\,\frac{1}{v} ,
}
\end{equation}
provided speed $v=v(R)$ and rotational diffusion coefficient $\Drot=\Drot(R)$ depend on distance $R$ but not on direction 
$\psi$.
Hence, $P^\steady(\R,\psi)$ is independent of $\Drot$ and inversely proportional to $v$, 
with isotropically distributed direction angles $\psi$.
This known fact can be derived also from the Fokker-Planck equation of ABP in Cartesian coordinates,
$\dot{P}(\x,\t,t) = 
- \partial_\x v(\x)\t P + \Drot(\x) \triangle_\t P$,
where $\triangle_\t$ denotes the spherical Laplacian. 

Of note, 
previous authors derived a one-dimensional phenomenological diffusion law
with position-dependent translational degrees of freedom
for ABP with position-dependent speed by averaging out directional degrees of freedom \cite{Sharma2017,Vuijk2018},
using Fox' colored noise approximation \cite{Fox1986}.
In short, directional persistence with persistence time $\tau_p$ 
is approximated as colored noise with same correlation time.
In the limit $\tau_p\rightarrow 0$,
a noise-induced drift term remains \cite{Vuijk2018}, 
which implies $P^\steady(\x)\sim 1/v(\x)$.
The fact that persistent random walks will penetrate a certain distance into a zone with high directional fluctuations 
before the distribution of their directions becomes isotropic
is implicit in these coarse-grained theories \cite{Sharma2017,Vuijk2018}
in the choice of stochastic calculus used to interpret position-dependent translational diffusion coefficients \cite{Lau2007}.

\subsection{Distribution of direction angles for persistent random walks crossing a boundary}
\label{sec:p_psi}

We compute the probability density $p(\psi)$ of direction angles $\psi$
at which ABPs will cross a spherical shell $S_i$ of radius $R_i$.
The Langevin equation Eqs.~(1,2) can be rewritten as a Fokker-Planck equation
$\dot{p}(R,\psi;t)=-\nabla\cdot\mathbf{J}$ 
with probability current $\mathbf{J}$
that satisfies $\mathbf{J}\cdot\e_R = v\cos\psi\,p(R,\psi;t)$.

We make the simplifying assumption that tangent directions are isotropically distributed, $p(R,\psi;t)\sim\sin\psi$. 
Thus,
the direction angles $\psi$ of trajectories $\R(t)$
crossing $S_i$ from the outside to the inside are distributed according to
$p(\psi)=\sin(2\psi)$ for $0\le\psi<\pi/2$ and $p(\psi)=0$ else,
whereas the direction angles of trajectories
crossing $S_i$ from the inside to the outside are distributed according to
$p(\psi)=\sin(-2\psi)$ for $\pi/2\le\psi\le\pi$ and $p(\psi)=0$ else.

\begin{figure}
\includegraphics[width=\linewidth]{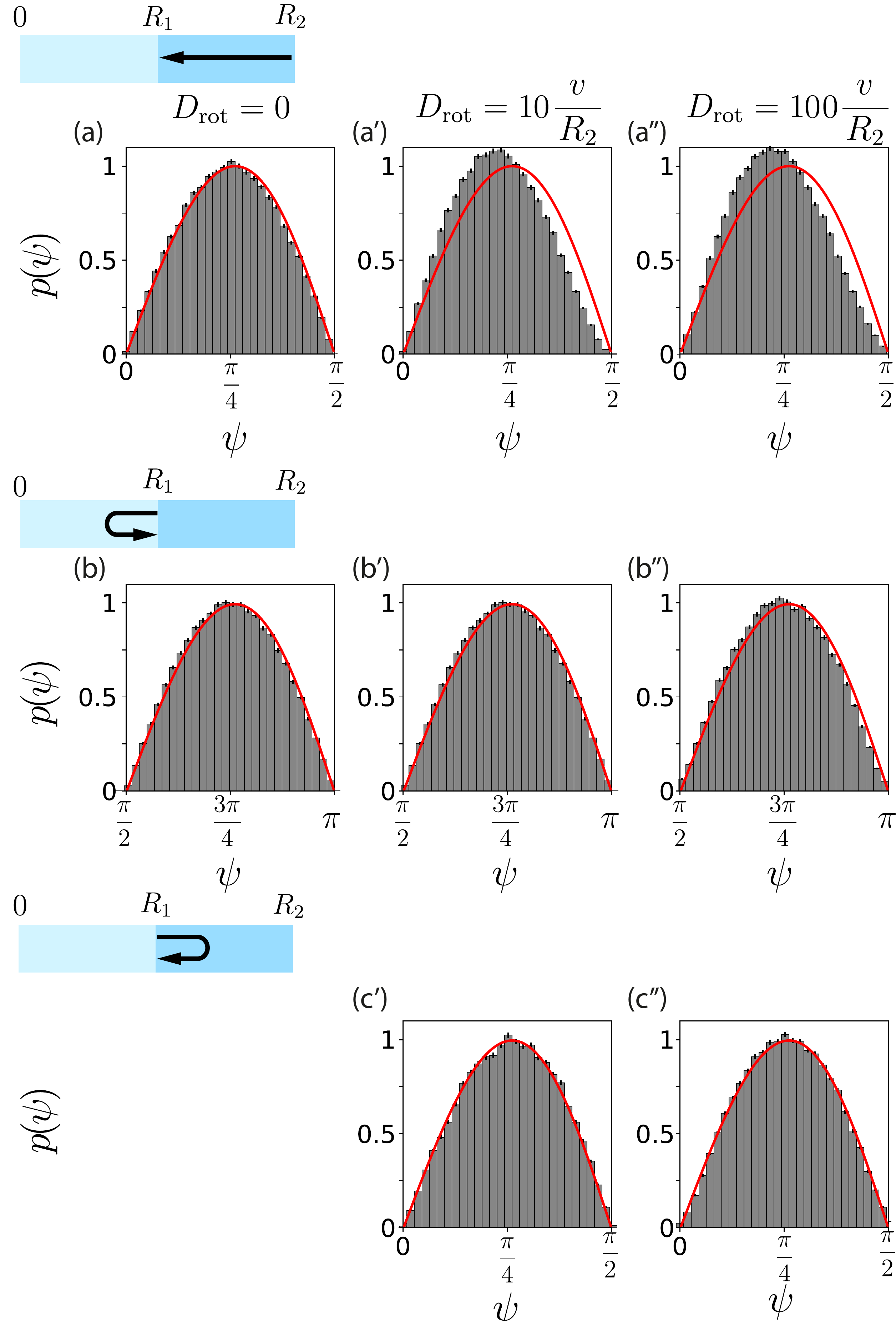}
\caption{
\textbf{Distribution of direction angles.} 
(a): 
Distribution $p(\psi)$ of direction angles $\psi$
(with $\t\cdot\e_R = \cos\psi$)
at the endpoint of simulated trajectories 
that started at $R_2$ with initial angle distributed according to $p(\psi)=\sin 2\psi$, $0\le\psi\le\pi/2$,
traversed zone 2 and ended at $R_1$
for the case of ballistic motion with $\Drot=0$. 
(a$'$): 
Same as panel (a), but for $\Drot=10\, v/R_2$. 
(a$''$): 
Same as panel (a), but for $\Drot=10^2\, v/R_2$. 
(b), (b$'$), (b$''$): 
Same as panels (a), (a$'$), (a$''$), 
but for trajectories that started at $R_1$ 
(with initial angle distributed according to $p(\psi)=\sin 2\psi$, $0\le\psi\le\pi/2$), 
traversed zone 1 and ended again at $R_1$. 
(c$'$), (c$''$): 
Same as panels (a$'$), (a$''$), 
but for trajectories that started at $R_1$ 
(with initial angle distributed according to $p(\psi)=-\sin 2\psi$, $\pi/2\le\psi\le\pi$), 
traversed zone 2 and ended again at $R_1$. 
Note that this is impossible for ballistic motion, hence no panel (c). 
Solid red line indicates approximate analytic solution, 
$p(\psi)=\sin(2\psi)$ for $0\le\psi\le\pi/2$ [panels (a), (a$'$), (a$''$), (c$'$), (c$''$)], and
$p(\psi)=\sin(-2\psi)$ for $\pi/2\le\psi\le\pi$ [panels (b), (b$'$), (b$''$)].
Parameters, see Fig.~2.
}
\label{fig_p_psi}
\end{figure}

The assumption of isotropically distributed directions may not be valid
if we restrict ourselves to specific initial conditions, 
e.g.\ trajectories that start at a distance $R_{i+1}$ and become absorbed at $R_i$.
Fig.~\ref{fig_p_psi} 
displays simulation results for $p(\psi)$ 
for different initial conditions, 
demonstrating that $p(\psi)=\sin(2\psi)$ still holds approximately.

We can find an analytical result for the case of ballistic trajectories,
showing that $p(\psi)$ converges to $\sin(2\psi)$ if the initial position is sufficiently far from $R_i$. 
We consider a spherical shell $S_i$ with radius $R_i$, and 
a current of ballistic trajectories starting at each position of a concentric spherical shell with radius $R_{i+1}$ 
with constant rate $j_0$ and isotropically distributed direction.
The corresponding flux as function of the direction angle $\psi_0$ at the start position reads
$j(\psi_0)=j_0\sin(\psi_0)/2$.
A ballistic trajectory will hit the sphere $S_i$ if and only if $\sin\psi_0\le\gamma$,
where $\gamma=R_i/R_{i+1}<1$.
For these trajectories, 
the direction angle $\psi_i$ at $S_i$ is related to $\psi_0$ by 
$\sin\psi_i=\sin\psi_0/\gamma$.
We can parameterize this sub-ensemble of ballistic trajectories by either $\psi_0$ or $\psi_i$. 
Using
$j(\psi_i)\sim j(\psi_0) (d\psi_0/d\psi_i)$, 
we obtain for the distribution of angles $p(\psi_i)$ of trajectories hitting $S_i$
due to $p(\psi_i)\sim j(\psi_i)$
\begin{equation}
p(\psi_i)=\frac{1}{2}\,\frac{\gamma^2}{1-\sqrt{1-\gamma^2}}\, \frac{\sin(2\psi_i)}{\sqrt{1-\gamma^2\sin^2\psi_i}}
\end{equation} 
for $0\le\psi_i<\pi/2$ and $p(\psi_i)=0$ else.
Thus, for $\gamma=1$, $p(\psi_i)=\sin(\psi_i)$,
while in the limit $\gamma\ll 1$ 
(corresponding to large zones with $R_0/R_1\ll 1$ and a limit of sparse targets), 
we recover the result for persistent random walks derived above, $p(\psi_i)=\sin(2\psi_i)$
for $0\le\psi_i<\pi/2$ and $p(\psi_i)=0$ else.

\subsection{Derivation of Eq.~(5)}

We consider a ballistic trajectory $\mathbf{r}(t)$ starting at a point $\r(t_0)$ on a sphere of radius $R_2$, 
whose tangent vector encloses an angle $\psi$ with the radial direction vector $\e_R$ at the start point $\r(t_0)$.
This trajectory will hit a sphere of radius $R_1$ with $R_1<R_2$ concentric with the first sphere 
if and only if $0\le\psi\le\psi_\mathrm{max}$, where $\sin\psi_\mathrm{max}=R_1/R_2$.
For a probability density $p(\psi)=\sin(2\psi)$ of initial direction angles with $0\le\psi\le\pi/2$, 
we obtain for the probability to hit the sphere of radius $R_1$
\begin{equation}
\int_0^{\psi_\mathrm{max}} \! d\psi \, p(\psi) = (R_1/R_2)^2.
\end{equation}

\subsection{Derivation of Eq.~(6)}

We provide additional details on the derivation of Eq.~(6) 
for the return probability $p_\mathrm{ret}$.

We first consider the limit $\lambda_2\ll 1$.
An ABP entering zone 2 from zone 1 at time $t=0$
will first continue moving in approximately radial direction 
before the tangent direction of its trajectory decorrelates.
If the direction angle $\psi_0$ of the tangent vector $\t(0)$ at $t=0$ 
is randomly distributed according to
$p(\psi_0)=\sin(-2\psi_0)$ for $\pi/2\le\psi_0\le\pi$
and $p(\psi_0)=0$ else, 
we have 
$\langle \t(t)\cdot\e_R(0) \rangle = -(2/3)\exp(-t/\tau_p)$ with $\tau_p=l_2/v$.
For times $t\gg\tau_p$, the ABP will exhibit isotropic random motion.
We introduce a cross-over time $t_0$ with $\tau_p\ll t_0\ll D_2 (d_2/v_0)^2$.

For the first dynamic phase defined by $t\leq t_0$,
we introduce the non-normalized probability density $p(x,t)$ to find an ABP at radial position $R_1+x$ at time $t$, 
where we impose absorbing boundary conditions at $R=R_1$.
The survival probability reads $Q(t)=\int_0^{d_2} dx\,p(x,t)$.
We are interested in the conditional expectation value
\begin{equation}
\langle x(t)\rangle=\int_0^{d_2} dx\,x p(x,t)/Q(t), 
\end{equation}
i.e., how deep the surviving ABPs have penetrated into zone 2. 
In the limit $l_2\ll R_1$, 
we may approximate the absorbing spherical shell at $R=R_1$ by a plane $H$ with surface normal $\n=\e_R(\R(0))$. 
We show that 
$\lim_{t\rightarrow\infty} \langle x(t)\rangle Q(t) = \alpha l_2$ with $\alpha=4/3$.

We first consider the problem without absorbing boundary conditions at $H$,
with corresponding probability density $p_0(x,t)$ to find an ABP at a distance $x$ from $H$ at time $t$, 
and an analogous normalized penetration depth 
$\alpha_0 = \lim_{t\rightarrow\infty} \int_0^\infty\! \langle x(t)\rangle_0/l_2 = \lim_{t\rightarrow\infty} \int_0^\infty\! dx\, x p_0(x,t)/l_2$.
Here, the expectation value $\langle\cdot\rangle_0$ averages over all persistent random walks, 
including those that have crossed $H$ at some time $t_1\le t$. 
By integrating the correlation function $\langle \t(t)\cdot\e_R(0)\rangle$ for $0\le t<\infty$, 
we show 
$\alpha_0 = 2/3$.
Specifically, let $\e_1=\t_0$ with $\t_0=\t(0)$, $\e_3=\t_0\times\n/|\t_0\times\n|$, $\e_2=\e_3\times\e_1$.
We write $\n=a_1\e_1+a_2\e_2$. 
Now, 
\begin{align*}
\langle \t(t)\cdot\n\rangle 
&= \langle a_1\,\t(t)\cdot\e_1\rangle + \langle a_2\,\t(t)\cdot\e_2\rangle \\
&= \langle a_1\rangle\,\langle\t(t)\cdot\e_1\rangle + \langle a_2\rangle\,\langle\t(t)\cdot\e_2\rangle \\
&= \frac{2}{3}\,\exp(-t/\tau_p) - \frac{1}{3}\, 0.
\end{align*}
Since $\R(t)=v_0\int_0^t\! dt'\,\t(t')$, 
we conclude 
$\alpha_0=(2/3)\,v_0 \tau_p=(2/3)\,l_2$.

Next, we argue 
$\alpha_0 = \alpha - \alpha_0$.
The expectation value in the definition of $\alpha_0$ can be decomposed into 
a contribution from the ABP that have never returned to $H$, equal to $\alpha$, and
a contribution from those ABP that returned to $H$, which yields $-\alpha_0$ in the long-time limit.
For a proof, 
consider the sub-ensemble of ABPs that have returned to $H$ at a time $t_1$.
The direction angles $\psi_1$ of their tangent vectors $\t(t_1)$ will be randomly distributed 
in a range $0\le\psi_1\le\pi/2$, with probability density given by $p(\psi_1)=\sin(2\psi_1)$.
Thus, this sub-ensemble corresponds to the original problem, but reflected at $H$ and starting at time $t_1$.
We now spell out this argument in detail. 
First, we split the probability density $p_0(x,\psi,t)$ as 
\begin{equation}
p_0(x,\psi,t)=p(x,\psi,t)+p_1(x,\psi,t), 
\end{equation}
where $p(x,\psi,t)$ is the unnormalized probability density of persistent random walks 
for the case of absorbing boundary conditions at $H$ introduced above
(corresponding to persistent random walks that have never returned to $H$ at any time before $t$), 
as well as a probability density $p_1(x,\psi,t)$ of persistent random walks 
that have crossed $H$ at some time $t_1$ with $t_1\le t$. 
When persistent random walks cross $H$ for the first time at some time $t_1$, 
their tangent $\t_1=\t(t_1)$ satisfies $\t_1\cdot\n>0$ 
with an angle $\psi_1$ enclosed by $\t_1$ and $\n$ that satisfies
$p(\psi_1)=\sin(2\psi_1)$ for $0\le\psi_1\le \pi/2$ and $p(\psi_1)=0$ else, 
see subsection `Distribution of direction angles'.
Thus, these persistent random walks correspond
to the same ensemble of persistent random walks considered in the definition of $\alpha_0$
after a mirror operation at $H$ and a time shift has been applied.
We conclude
$\alpha_1 = \lim_{t\rightarrow\infty} \int\! dx\,d\psi\, x p_1(x,\psi,t) = -\alpha_0$.
Consequently, 
$\alpha_0 = \alpha + \alpha_1 = \alpha - \alpha_0$,
hence $\alpha=4/3$.
Here, we used that ABPs will eventually return to $H$ with probability one,
with a survival probability that decays for $t\gg \tau_p$
like the survival probability of diffusive particles, $Q(t)\sim (t/\tau_p)^{-1/2}$, 
see section `Asymptotic scaling of survival probability'.

Interestingly, 
it follows that the conditional expectation value
$\langle x(t)\rangle/Q(t)$ 
diverges as $(t/\tau_p)^{1/2}$, 
i.e.,\ while fewer and fewer ABP survive, 
their mean distance from $H$ increases with time 
such that $\langle x(t)\rangle_s/l_2$ converges to the definite value $\alpha=4/3$.

We now address the second dynamic phase defined by $t\ge t_0$, 
and calculate $p_\mathrm{ret}$.
Those ABPs that have not been absorbed at $R_1$ before $t_0$
will likely be found at a distance $x\gg l_2$ from $R_1$, 
and we may approximate these as diffusive particles.
The probability that a diffusive particle reaches $R_2$ if released at a radial position $R$
between two absorbing spherical shells of radii $R_1$ and $R_2$ reads \cite{Berg1977}
\begin{equation}
p(R)=\frac{R_2}{R}\,\frac{R-R_1}{R_2-R_1}. 
\end{equation}
We choose
$R=R_1+\langle x(t_0)\rangle \approx R_1+\alpha l_2/Q(t_0)$
and obtain an asymptotic result for 
$q=1-p_\mathrm{ret}$
as
\begin{equation}
\label{eq:q_lim}
q \approx 
Q(t_0)p[R(t_0)] 
 \approx \alpha\frac{R_2}{R_1}\lambda_2,
\end{equation}
which is valid for $\lambda_2\ll (\tau_p/t_0)^{1/2}$.
We can extend this asymptotic expression to the entire range $0<\lambda_2<\infty$ by interpolating 
with the limit value 
$q^\ballistic=\lim_{\lambda_2\rightarrow\infty}1-p_\mathrm{ret}=1$, 
using the simple ansatz of a saturation curve 
\begin{equation}
\label{eq:q_interp}
q \approx 
\frac{q^\ballistic\,\gamma\lambda_2}{ q^\ballistic+\gamma\lambda_2 }
\end{equation}
with 
$\gamma = {\partial q / \partial\lambda_2}_{|\lambda_2=0}$
computed using Eq.~(\ref{eq:q_lim}),
i.e., we match the initial slope at $\lambda_2=0$ in Eq.~(\ref{eq:q_interp}) and Eq.~(\ref{eq:q_lim}).
We conclude Eq.~(6)
\begin{equation}
\nonumber
\boxed{
p_\mathrm{ret} \approx 
\left( 1 + \alpha \lambda_2\,R_2/R_1 \right)^{-1}, \quad \alpha=4/3.
}
\end{equation}
An analogous derivation yields
\begin{equation}
p_i \approx \frac{p_i^\ballistic\,\alpha\lambda_i\, R_{i-1}}{R_i p_i^\ballistic+\alpha\lambda_i R_{i-1}}
\end{equation}
for $i=1,2$.

\begin{figure}
\includegraphics[width=\linewidth]{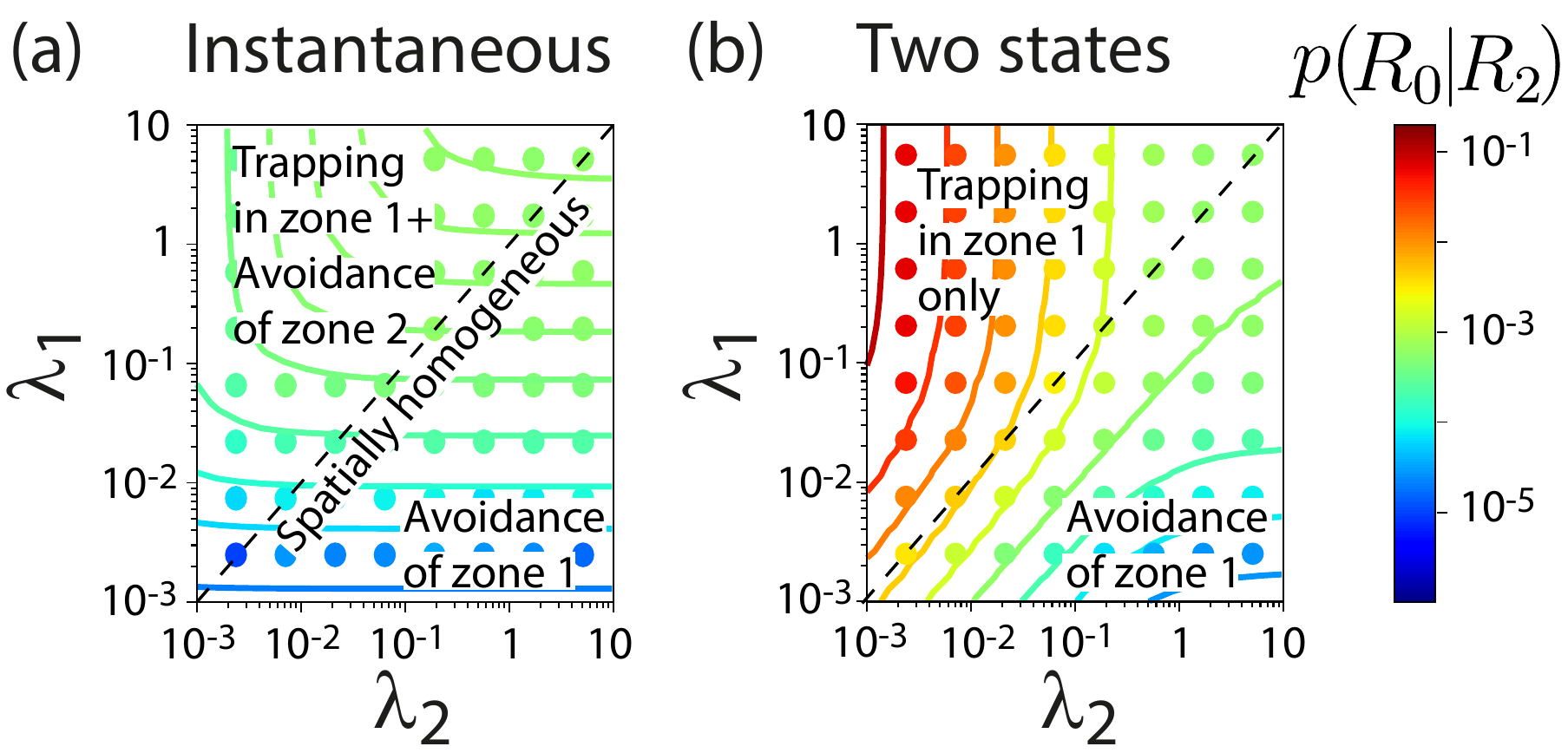}
\caption{
\textbf{Target-finding probability as function of zone-dependent persistence length.}
(a): 
Probability $p(R_0|R_2)$ for ABP with instantaneous chemokinesis starting at distance $R_2$ to hit a target of radius $R_0$ at the origin
as function of normalized persistence lengths
$\lambda_i=l_i/d_i$ 
(with $l_i=v/(2D_i)$, $d_i=R_i-R_{i-1}$) 
in zones $i=1,2$,
analogous to Fig.~\ref{figure1}(b).
The success probability is maximal for ballistic motion with $\lambda_1,\lambda_2\gg 1$.
(b):
Same as panel (a), but for ABP with two-state chemokinesis as in Fig.~\ref{figure1}(c). 
The probability $p(R_0|R_2)$ becomes maximal for $\lambda_1\gg 1$, $\lambda_2\ll 1$.
In this regime, two-state chemokinesis can harness
inward scattering of outgoing trajectories (`trapping in zone 1'), 
while reducing outward scattering of ingoing trajectories (`avoidance of zone 2').
Dots: numerical simulations; 
solid lines: analytic theory, Eqs.~(\ref{eq:pin}), (\ref{eq:prefl}).
Parameters, see Fig.~\ref{figure1}, unless stated otherwise.
}
\label{fig:p_success}
\end{figure}

The return probability $p_\mathrm{ret}$ and the zone-crossing probabilities $p_1$, and $p_2$ 
allow to compute the probability $p(R_0|R_2)$ by Eq.~(\ref{eq:pin}). 
For a comparison of analytic theory and numerical simulations, see also Fig.~\ref{fig:p_success}.

\subsection{Asymptotic scaling of survival probability $Q(t)$}
\label{sec:q_asymptotic}

In the main text, we introduced the `normalized penetration depth' $\alpha$ of persistent random walks 
with persistence length $l_p$ starting at time $t=0$ at a plane boundary $H$ with absorbing boundary conditions at $H$, 
there defined as
\begin{equation}
\label{eq:alpha}
\alpha = \frac{1}{l_p}\,\lim_{t\rightarrow\infty} \langle x \rangle / Q(t).
\end{equation} 
We provide a heuristic argument why the limit $\alpha$ in Eq.~(\ref{eq:alpha}) exists and is finite. 

We choose some $t_0>0$ such that $\langle x(t_0)\rangle \gg l_p$. 
We replace the persistent random walk $\R(t)$
by a diffusive trajectory with translational diffusion coefficient $D=l_P v_0/3$
for $t\ge t_0$.
Thus, both the persistent random walk 
and the diffusive trajectory will have identical statistical properties 
on length scales large compared to $l_p$. 

For a diffusive trajectory starting at time $t_0$ at an initial distance $x_0$ from $H$, 
with absorbing boundary conditions at $H$,
the unnormalized probability density $\widehat{p}(x,t_0+\tau|x_0,t_0)$ of surviving trajectories reads
\begin{equation}
\label{eq:phat}
\widehat{p}(x,t_0+\tau|x_0,t_0) = N(x;x_0,2 D \tau) - N(x;-x_0,2 D\tau),
\end{equation}
where $N(x,\mu,\sigma^2) = (2\pi\sigma^2)^{-1/2} \exp[-(x-\mu)^2/(2\sigma^2) ]$ 
denotes the normal distribution with mean $\mu$ and variance $\sigma^2$.
We decorate variables by a hat to indicate quantities related to diffusive trajectories 
in contrast to persistent random walks.  

The survival probability of diffusive trajectories is given by
\begin{equation}
\widehat{Q}(t_0+\tau|x_0,t_0) = \mathrm{Erf}\left( \sqrt{ \frac{x_0^2}{4 D \tau} } \right).
\end{equation}
Note $\widehat{Q}(t_0+\tau)\sim \sqrt{x_0^2/(D \tau)}$ for $\tau\gg \sqrt{x_0^2/D}$. 
For the first moment of $\widehat{p}(x,t_0+\tau|x_0,t_0)$, we find
\begin{equation}
\label{eq:x0}
\int_0^\infty\! dx\, x\,\widehat{p}(x,t_0+\tau|x_0,t_0) = x_0. 
\end{equation}
Eq.~(\ref{eq:x0}) follows directly from the analytical solution Eq.~(\ref{eq:phat}) 
for the probability density $\widehat{p}(x,t_0+\tau|x_0,t_0)$.

Eq.~(\ref{eq:x0}) can be deduced also by a direct argument:
in the absence of absorbing boundary conditions at $H$, we have
\begin{equation}
\langle x(t_0+\tau)|x_0,t_0\rangle_0
= \int_{-\infty}^\infty dx\, x \widehat{p}_0(x,t_0+\tau|x_0,t_0) = x_0, 
\end{equation}
where 
$\widehat{p}_0(x_0,t_0+\tau|x_0,t_0) = N(x;x_0,2 D \tau)$ 
is the probability density without absorbing boundary. 
If we now restrict the computation of the mean to those realizations
that passed through $x=0$ at a time $t_1$ with $t_0<t_1\le t$, we have
$\langle x(t)|x(t_1)=0,x(t_0)=x_0\rangle_0=0$ 
by the Markov property of random walks. 
We conclude 
\begin{equation}
\int_0^\infty dx\, x \widehat{p}(x,t_0+\tau|x_0,t_0) = 
\int_{-\infty}^\infty dx\, x \widehat{p}_0(x,t_0+\tau|x_0,t_0), 
\end{equation}
i.e.,
the expectation value of $x(t)$ does not change
if those realization that returned to $x=0$ at some time $t_1$ with $0<t_1<t$ are not included in the integral.

As a corollary, 
the conditional mean value
$\langle x(t_0+\tau)\rangle$
for diffusive trajectories starting with $x(t_0)=x_0$ at $t=t_0$
that have not crossed $x=0$ obeys
\begin{equation}
\langle x(t_0+\tau)\rangle=x_0/\widehat{Q}(t_0+\tau)\sim \tau^{1/2}. 
\end{equation}

\subsection{Numerical simulations for the normalized penetration depth $\alpha$}

\begin{figure}[h]
\begin{center}
\includegraphics[width=1\linewidth]{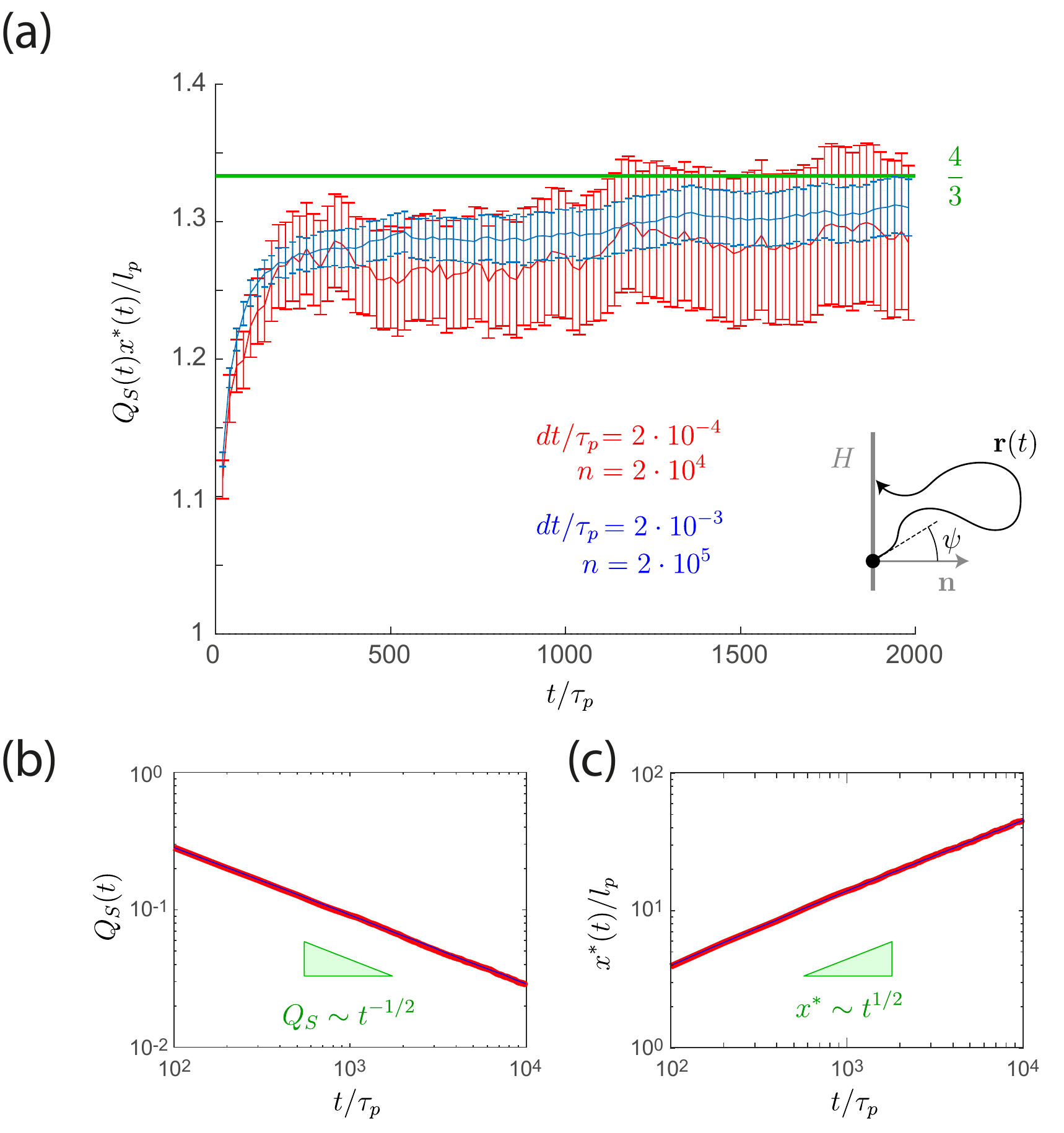}
\end{center}
\caption{
\textbf{Numerical simulations for the normalized penetration depth $\alpha$.}
(a):
Simulation results for $\langle x(t)\rangle Q(t)$ as function of normalized time $t/\tau_p$, 
illustrating the limit in Eq.~(\ref{eq:alpha}).
Shown are results for time steps $dt/\tau_p=2\cdot 10^{-3}$ (blue, $n=2\cdot 10^5$ trajectories), and 
$dt/\tau_p=2\cdot 10^{-4}$ (red, $n=2\cdot 10^4$ trajectories).
Error bars denote standard error of the mean determined by bootstrapping with replacement.
Green line: approximate analytical result $\alpha=4/3$.
Inset: schematic of a persistent random walk $\R(t)$ 
starting at a plane $H$ with surface normal $\mathbf{n}$ at time $t=0$, 
and ultimately being absorbed at $H$ again at a later time $t_1$,
with initial tangent $\t(0)$ enclosing an angle $\psi$ with $\mathbf{n}$.
(b):
Survival probability $Q(t)$ as function of $t/\tau_p$
(blue: $dt/\tau_p=2\cdot 10^{-3}$, red: $dt/\tau_p=2\cdot 10^{-4}$). 
Simulation results confirm the asymptotic scaling $Q(t)\sim t^{-1/2}$. 
(c):
Conditional mean distance $\langle x(t)\rangle$ from the plane $H$ as function of $t/\tau_p$
[colors as in panel (b)]. 
Simulations results confirm the asymptotic scaling $\langle x(t)\rangle \sim t^{1/2}$. 
\label{fig:alpha}
}
\end{figure}

We compared the analytical approximation of the constant $\alpha$ to numerical simulations, 
see Fig.~\ref{fig:alpha}.
These numerical simulations provided the estimate $\alpha\approx 1.310\pm 0.021$ 
(mean$\pm$s.e.m.).
The standard error of the mean was determined by bootstrapping with replacement. 
This estimate can be considered a lower bound since a finite simulation time was used.
We employed an Euler scheme using matrix exponentials for propagation of the Frenet-Serret frame;
CPU time $250$ hours on a standard PC.
Analogous simulations using a smaller time step gave consistent results.




\subsection{Derivation of Eq.~(7)}

We provide additional details for the derivation of Eq.~(7).
This ordinary differential equation represents a backward equation for the penetration probability $p(R_0|R')$ 
that an adaptive ABP starting at distance $R'$ will reach distance $R_0$ before returning to $R'$, 
where the ABP performs instantaneous chemokinesis with rotational diffusion coefficient $\Drot=\Drot(R)$,
where $\Drot(R)$ is some arbitrary function of radial distance $R$.
In the definition of $p(R_0|R')$, we assume a random inward pointing initial direction angle $\psi$, 
distributed according $p(\psi)=\sin(2\psi)$ for $0\le\psi\le\pi/2$ and $p(\psi)=0$ else.

\begin{figure}[h]
\begin{center}
\includegraphics[width=1\linewidth]{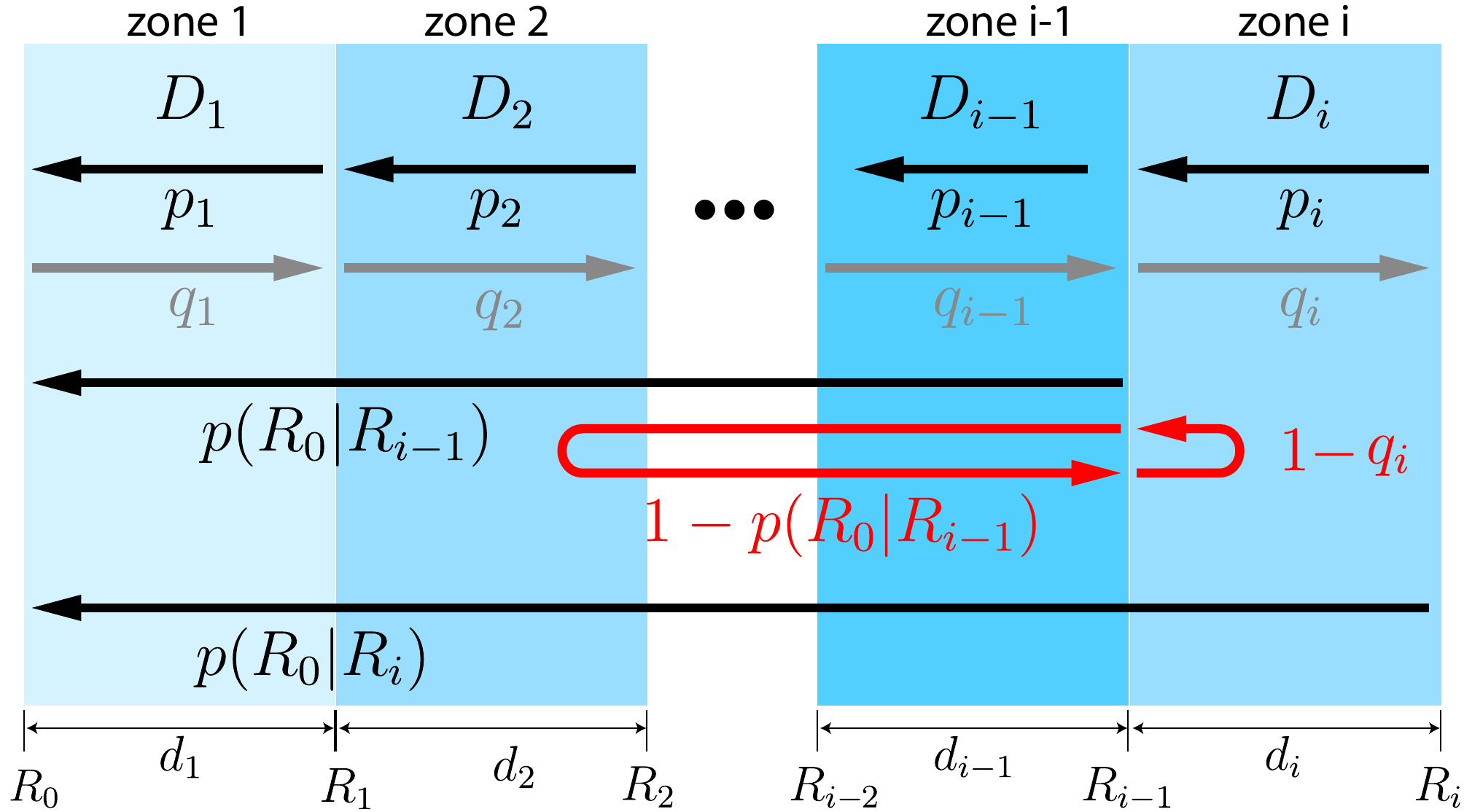}
\end{center}
\caption{
\label{fig:zones}
\textbf{Notation used in derivation of Eq.~(7).}
We consider spatial zones bounded by $R_{i-1}\le R\le R_i$, $i=1,\ldots,n$,
where the rotational diffusion coefficient $\Drot$ takes a constant value $D_i$ in zone $i$.
We introduce splitting probabilities $p_i$ and $q_i$ to `cross' zone $i$ in inward or outward direction, respectively, 
see Eq.~(\ref{eq:pi_def}) and Eq.~(\ref{eq:qi_def}).
We derive a recursion relation for $p(R_0|R_i)$ 
in terms of $p(R_0|R_{i-1})$, $p_i$, and $q_i$, see Eq.~(\ref{eq:pi_back}).
}
\end{figure}

We first consider a case of $n$ zones bounded by spheres concentric with the origin, 
$R_{i-1}\le R< R_i$, $i=1,\ldots,n$, 
where we assume that $\Drot$ takes the constant value $D_i$ in zone $i$, see Fig.~\ref{fig:zones}.
We introduce short-hand 
$l_i=v/(2 D_i)$ for the corresponding persistence length, and
$\lambda_i = l_i / d_i$ for the ratio of $l_i$ and the width $d_i=R_i-R_{i-1}$ of zone $i$.

Let us consider zone $i$.
We impose absorbing boundary conditions at $R=R_{i-1}$ and $R=R_i$ and 
define the splitting probabilities
\begin{equation}
\label{eq:pi_def}
p_i = \pi_{R_{i-1},R_i} (R_{i-1}\,|\,R_i,\leftarrow ), 
\text{ and }
1-p_i
\end{equation}
that an ABP starting at distance $R=R_i$ from the origin
will reach either the boundary at $R=R_{i-1}$ or the other boundary at $R=R_i$ first, respectively.
Here, the arrow `$\leftarrow$' shall indicate a random inward pointing initial direction angles,
distributed according to $p(\psi)=\sin(2\psi)$ for $0\le\psi\le\pi/2$ and $p(\psi)=0$ else.

We define analogous splitting probabilities for the opposite direction, 
i.e., the splitting probabilities
\begin{equation}
\label{eq:qi_def}
q_i = \pi_{R_{i-1},R_i} (R_i\,|\,R_{i-1},\rightarrow ),
\text{ and }
1-q_i
\end{equation}
that an ABP starting at distance $R=R_{i-1}$ from the origin
will reach either an absorbing boundary at $R=R_i$ or $R=R_{i-1}$ first, respectively.
Here, the arrow `$\rightarrow$' indicates a random outward pointing initial direction angles
[distributed according to $p(\psi)=\sin(-2\psi)$ for $\pi/2\le\psi\le\pi$ and $p(\psi)=0$ else].
Note $p_\mathrm{ret}=1-q_2$.

We quote the previous results for $p_i$ and $q_i$, $i=1,\ldots,n$ (derived above for $i=2$)
\begin{align}
\label{eq:pq}
p_i &= \frac{ \alpha \lambda_i R_{i-1}/R_i }{ 1 + \alpha \lambda_i R_i/R_{i-1} } , \\
q_i &= \frac{ \alpha \lambda_i R_i/R_{i-1} }{ 1 + \alpha \lambda_i R_i/R_{i-1} } .
\end{align}
Now, 
we consider zones $1,2,\ldots,i$ together, 
and impose absorbing boundary conditions only at $R_0$ and $R_i$.
We introduce the splitting probability
\begin{equation}
p(R_0|R_i) = \pi_{R_0,R_i} (R_0\,|\,R_i,\leftarrow )
\end{equation}
that an ABP starting at $R_i$ will reach $R_0$ instead of returning to $R_i$, 
where we assume a random inward pointing initial direction angle
(distributed according $p(\psi)=\sin(2\psi)$ for $0\le\psi\le\pi/2$).
Analogously, we introduce the splitting probability for the opposite direction
\begin{equation}
q(R_i|R_0) = \pi_{R_0,R_i} (R_i\,|\,R_0,\rightarrow )
\end{equation}
that an ABP starting at $R_0$ will reach $R_i$ instead of returning to $R_0$, 
where we assume random outward pointing initial direction angle
[distributed according $p(\psi)=\sin(-2\psi)$ for $\pi/2\le\psi\le\pi$ and $p(\psi)=0$ else].

We will derive recursion relations for $p(R_0|R_i)$ and $q(R_i|R_0)$.

\paragraph{Backward equation.}
Eq.~(4) generalizes in a straight-forward manner to 
\begin{equation}
\label{eq:pi_back}
p(R_0 | R_i ) 
\approx 
\sum_{k=0}^\infty p_i\,[(1-p(R_0|R_{i-1}) (1-q_i)]^k\, p(R_0|R_{i-1}).
\end{equation}
By symmetry, we have a similar equation for $q(R_i|R_0)$
\begin{equation}
q(R_i | R_0 ) 
\approx 
\sum_{k=0}^\infty q(R_{i-1}|R_0)\,[(1-q_i)(1-p(R_0|R_{i-1}))]^k\,q_i.
\end{equation}

In the limit $d_i\ll R_i$ with $\lambda_i\gg 1$, 
we can expand Eq.~(\ref{eq:pi_back}) in powers of $d_i$, using Eq.~(\ref{eq:pq})
\begin{align}
p(R_0 | R_i ) 
& \approx 
\frac{ p(R_0|R_{i-1})\,p_i }{1- [1-p(R_0|R_{i-1})](1-q_i)} \\
& =
p(R_0|R_{i-1}) \notag \\
& \phantom{=} 
- \left[ 
\frac{2}{R_i} p(R_0|R_{i-1}) 
+ \frac{1}{\alpha l_i} p(R_0|R_{i-1})^2
\right] d_i \notag \\
& \phantom{=} + \mathcal{O}(d_i^2) .
\end{align}
The continuum limit $d\rightarrow 0$ gives the nonlinear differential equation
\begin{equation}
\label{eq:p_ode_back}
\boxed{
\frac{\partial}{\partial R} p(R_0|R) 
=
- \frac{2}{R}\, p(R_0|R) - \frac{1}{\alpha l_p(R)}\, p(R_0|R)^2
}
\end{equation}
with position-dependent persistence length $l_p(R)=v/[2\Drot(R)]$.

Similarly, we find for $q(R_i|R_0)$
\begin{align}
q(R_i | R_0) 
& \approx 
\frac{ q(R_{i-1}|R_0)\,q_i }{ 1- [(1-q_i)(1-p(R_0|R_{i-1}))] } \\
& =
q(R_{i-1}|R_0) \notag \\
& \phantom{=} 
- \frac{1}{\alpha l_i} \, p(R_0|R_{i-1})\, q(R_{i-1}|R_0) d_i + \mathcal{O}(d_i^2) .
\end{align}
The continuum limit gives the differential equation
\begin{equation}
\label{eq:q_ode_back}
\frac{\partial}{\partial R} q(R|R_0) 
=
-\frac{1}{\alpha l_p(R)}\, p(R_0|R)\,q(R|R_0).
\end{equation}
We have initial conditions $p(R_0|R_0)=1$ and $q(R_0|R_0)=1$.
From Eq.~(\ref{eq:p_ode_back}), we obtain the analytical solution
\begin{align}
\label{eq:p_ana}
p(R_0|R')  & = 
\frac{ p^\ballistic(R_0|R') }
{
1 + 
	\int_{R_0}^{R'} dR \, 
		p^\ballistic(R_0|R) /( \alpha l_p(R) )  
}. 
\end{align}
Similarly, from Eq.~(\ref{eq:q_ode_back}), we obtain the analytical solution
\begin{align}
\label{eq:q_ana}
q(R'|R_0)  
&= 
\exp\left( 
	-\int_{R_0}^{R'} \! dR\,\,
		\frac{ p(R_0|R) }{ \alpha l_p(R) } 
\right) \notag, \\
&\stackrel{!}{=} p(R_0|R') / p^\ballistic(R_0|R') .
\end{align}
Here, we used the general relation
\begin{equation}
\label{eq:pq_sym}
\boxed{
p(R_0|R) = q(R|R_0)\,p^\ballistic (R_0|R) .
}
\end{equation}
A proof of this relation is obtained as follows:
let 
\begin{equation}
\theta(R) = p(R_0|R) / q(R|R_0). 
\end{equation}
From the differential equations for $p(R_0|R)$ and $q(R|R_0)$,
Eq.~(\ref{eq:p_ode_back}) and Eq.~(\ref{eq:q_ode_back}), 
we obtain an equation for $d\theta/dR$, 
where the terms dependent on $l_p$ cancel
\begin{align*}
\label{eq:theta_ode}
\frac{d}{dR} \theta 
&= 
  \frac{1}{ q(R|R_0) } \, \frac{\partial}{\partial R} p(R_0|R)
- \frac{p(R_0|R)}{q(R|R_0)^2} \, \frac{\partial}{\partial R} q(R_0|R) \notag \\
&=
-\frac{2}{R}\theta 
- \frac{1}{ \alpha l_p(R) } \left[ \frac{p(R_0|R)^2}{q(R|R_0)} - \frac{p(R_0|R)^2}{q(R|R_0)} \right] \notag \\
&=
-\frac{2}{R}\theta .
\end{align*}
Since $\theta(R_0)=1$, this equation yields $\theta(R) = p^\ballistic(R_0|R)$.
Eq.~(\ref{eq:pq_sym}) can be interpreted also as a corollary of time-reversal symmetry of Eqs.~(1,2) in the spirit of \cite{Benichou2005b}:
if we start with a homogeneous distribution of initial conditions on the union of spheres at $R_0$ and $R$
(with random outward pointing initial direction at $R_0$ and random inward pointing initial direction at $R$), 
then the stochastic dynamics of ABP will map this distribution onto a homogeneous distribution of end-points, 
only with direction of tangents reversed, to very good approximation.
Yet, for each trajectory, the time-reversed trajectory is equally probable; hence
$4\pi R_0^2\, q(R|R_0) = 4\pi R^2\, p(R_0|R)$.

\paragraph{Forward equation.}
In a similar fashion, we can derive a forward equation for $p(R|R_n)$.
Eq.~(4) generalizes to 
\begin{equation}
\label{eq:pi}
p(R_{i-1} | R_n ) 
\approx 
\sum_{k=0}^\infty p(R_i|R_n)\,[(1-p_i) (1-q(R_n|R_i))]^k\, p_i.
\end{equation}
By symmetry, we have a similar equation for $q(R_n|R_{i-1})$
\begin{equation}
q(R_n | R_{i-1}) 
\approx 
\sum_{k=0}^\infty q_i \, [(1-q(R_n|R_i))(1-p_i)]^k \, q(R_n|R_i) .
\end{equation}

In the limit $d_i\ll R_i$ with $\lambda_i\gg 1$, 
we can again expand Eq.~(\ref{eq:pi}) in powers of $d_i$, using Eq.~(\ref{eq:pq})
\begin{align}
p(R_{i-1} | R_n ) 
& \approx 
\frac{ p(R_i|R_n)\, p_i }{1- (1-p_i)(1-q(R_n|R_i))} \\
& =
p(R_i|R_n) \notag \\
& \phantom{=}
- \left[ \frac{2}{R_{i-1}} + \frac{1}{\alpha l_i} \right]\, p(R_i|R_n)\, q(R_n|R_i) d_i \notag \\
& \phantom{=} 
+ \mathcal{O}(d_i^2) .
\end{align}
The continuum limit $d\rightarrow 0$ gives the differential equation
\begin{equation}
\frac{\partial}{\partial R} p(R|R_n) 
=
\left[ \frac{2}{R} + \frac{1}{\alpha l_p(R)} \right] p(R|R_n)\, q(R_n|R)
\end{equation}
with position-dependent persistence length $l_p(R)=v/[2\Drot(R)]$ as above.
Similarly, we find for $q(R_n|R_{i-1})$
\begin{align}
q(R_n | R_{i-1}) 
& \approx 
\frac{ q_i \, q(R_n|R_i) }{1- [1-q(R_n|R_i)](1-p_i)} \\
& =
q(R_n|R_i) \notag \\
& \phantom{=}
+ \frac{2}{R_{i-1}}\, q(R_n|R_i) [1-q(R_n|R_i)]\, d_i \notag \\
& \phantom{=}
- \frac{1}{\alpha l_i} q(R_n|R_i)^2\, d_i + \mathcal{O}(d_i^2) .
\end{align}
The continuum limit gives the differential equation
\begin{align}
\frac{\partial}{\partial R} q(R_n|R) 
&=
\frac{2}{R}\, q(R_n|R)\,[1-q(R_n|R)] \notag \\
&\phantom{=}
- \frac{1}{\alpha l_p(R)} q(R_n|R)^2.
\end{align}

Using the symmetry relation between $p$ and $q$, Eq.~(\ref{eq:pq_sym}), 
we can derive an alternative forward equation for $p(R|R_n)$ that depends only on $p(R|R_n)$
\begin{align}
\boxed{
\frac{\partial}{\partial R} p(R|R_n) 
= 
\left( \frac{2}{R} + \frac{1}{\alpha l_p(R)} \right) 
\frac{ p(R|R_n)^2 }{ p^\ballistic(R|R_n) }.
}
\end{align}
The initial conditions are $p(R_n|R_n)=1$ and $q(R_n|R_n)=1$.

\paragraph{Forward and backward equation are equivalent.}
Forward and backward equation are mathematically equivalent.
This can be shown by inserting the formal solution, Eq.~(\ref{eq:p_ana}) and Eq.~(\ref{eq:q_ana}),
and using the symmetry relation, Eq.~(\ref{eq:pq_sym}).

For constant $\Drot(R)=D_1$, we recover the previous result for $p_1$.
In the limit of ballistic motion ($l_p\rightarrow\infty$), 
we recover $p\rightarrow p^\ballistic$, $q\rightarrow q^\ballistic$.

Our results include the case of zones bounded by parallel planes, corresponding to the limit $R_0\rightarrow \infty$.
We thus obtain equations for the case of a rotational diffusion coefficient that depends only on one spatial coordinate, 
$\Drot=\Drot(x)$ with $x=R-R_0$.
In this case, 
$p(R_0|R)=q(R|R_0)$ and the equations for $p$ and $q$ are symmetric, reflecting the mirror-symmetry of the problem's geometry.

\paragraph{Two-state chemokinesis ${\mathrm{(S)}}$.}
From the results for instantaneous chemokinesis, 
we can immediately infer splitting probabilities for a two-state agent.
We consider a two-state agent that initially moves ballistically,
starting at $R=R_2$ with random inward pointing initial direction angle
[distributed according to $p(\psi)=\sin(2\psi)$ for $0\le\psi\le\pi/2$ and $p(\psi)=0$ else]. 
Upon first reaching $R=R_1$, this agent shall permanently switch to 
instantaneous chemokinesis with position-dependent rotational diffusion coefficient $\Drot(R)$ as above.
The probability $p_{\mathrm{(S)}}(R_0|R_2)$ for the two-state agent 
to reach $R_0$ before returning to $R_2$ reads
\begin{equation} 
\boxed{
p_{\mathrm{(S)}}(R_0|R_2) \approx p(R_0|R_2)\, \frac{p^\ballistic(R_1|R_2) }{ p(R_1|R_2) }.
}
\end{equation}
The proof relies on the approximate product rule
\begin{align}
\pi_{R_0,R_2} (R_0 | R_2,\leftarrow ) 
& \approx 
\pi_{R_0,R_2} (R_0 | R_1,\leftarrow ) \, \cdot \notag \\
& \phantom{\approx} \ 
\pi_{R_0,R_2} (R_1 | R_2,\leftarrow ). 
\end{align}
Here, we used the fact that the distribution of orientation angles $\psi$ of agents arriving at $R_1$ 
will follow the stereotypic distribution $p(\psi)=\sin(2\psi)$, $0\le\psi\le\pi/2$, to very good approximation. 
Thus,
\begin{align}
p(R_0|R_2)  & = \pi_{R_0,R_2} (R_0|R_1,\leftarrow) \, p(R_1|R_2), \\
p_{\mathrm{(S)}}(R_0|R_2) & = \pi_{R_0,R_2} (R_0|R_1,\leftarrow) \, p^\ballistic(R_1|R_2),
\end{align}
and the assertion follows.

\paragraph{Arbitrary initial position.}
The results above generalize to the case where the adaptive ABP is released at an intermediate position between the absorbing boundaries.
Obviously, 
the end distance can always be chosen as one of the absorbing boundaries;
hence, for $i\le j$ 
\begin{align}
\pi_{R_i,R_n} (R_j\,|\,R_n,\leftarrow)  &= \pi_{R_j,R_n} (R_j\,|\,R_n,\leftarrow), \notag \\
\pi_{R_0,R_j} (R_i\,|\,R_0,\rightarrow) &= \pi_{R_0,R_i} (R_i\,|\,R_0,\rightarrow). 
\end{align}

Analogously to Eq.~(\ref{eq:pin}), 
we have a geometric series representation
\begin{multline}
\pi_{R_0,R_2} (R_0|R_1,\leftarrow) \\
 = p(R_0|R_1) \sum_{k=0}^\infty [(1-p(R_0|R_1))(1-q(R_2|R_1))]^k .
\end{multline}
Similarly, 
\begin{equation}
\pi_{R_0,R_2} (R_0|R_1,\rightarrow) = [1-q(R_2|R_1)]\,\pi_{R_0,R_2} (R_0|R_1,\leftarrow).
\end{equation}
Thus, it suffices to know $p(R_0|R_1)$ and $q(R_2|R_1)$.

\paragraph{Isotropic initial directions.}
To facilitate comparison with previous work, we also introduce
the probability $\pi_{R_0,R_2}(R_0|R_1,\leftrightarrow)$ to reach $R_0$ for an ABP starting at $R_1$
with isotropically distributed initial direction 
(i.e., $p(\psi)=(1/2)\,\sin\psi$ for $0\le\psi\le\pi$, indicated by `$\leftrightarrow$')
for absorbing boundary conditions at $R_0$ and $R_2$ (as considered e.g.\ in \cite{Vuijk2018}). 
We first consider the case of ballistic motion.
Analogous to Eq.~(5), we note the success probability for the case of ballistic motion 
for the different distribution of random initial directions
\begin{align}
p^\ballistic_{R_0,R_2}(R_0|R_1,\leftarrow) &= (R_0/R_1)^2, \\
p^\ballistic_{R_0,R_2}(R_0|R_1,\rightarrow) &= 0, \\
p^\ballistic_{R_0,R_2}(R_0|R_1,\leftrightarrow) &=  \frac{1}{2} \left( 1 - \sqrt{1 - \left( \frac{R_0}{R_1} \right)^2 } \right).
\end{align}
Note
$p^\ballistic_{R_0,R_2}(R_0|R_1,\leftrightarrow) = p^\ballistic_{R_0,R_2}(R_0|R_1,\leftarrow)/4$ for $R_1{\gg} R_0$, and
$p^\ballistic_{R_0,R_2}(R_0|R_1,\leftrightarrow){=}p^\ballistic_{R_0,R_2}(R_0|R_1,\leftarrow)/2$ for $R_1{\searrow} R_0$.

\paragraph{Comparison to Fox' colored noise approximation.}
The probability $\pi_{R_0,R_2}(R_0|R_1,\leftrightarrow)$ to reach $R_0$ for an ABP starting at $R_1$
with isotropically distributed initial direction (indicated by `$\leftrightarrow$')
for absorbing boundary conditions at $R_0$ and $R_2$ (as considered e.g.\ in \cite{Vuijk2018}) 
can be approximated as a weighted sum of the respective probabilities for ABP with 
inward and outward pointing initial directions, respectively
\begin{multline}
\label{eq:new_result}
\pi_{R_0,R_2}(R_0|R_1,\leftrightarrow) \\
\approx 
\beta\, \pi_{R_0,R_2} (R_0|R_1,\leftarrow) + (1-\beta)\, \pi_{R_0,R_2} (R_0|R_1,\rightarrow).
\end{multline}
Here, the interpolation coefficient $\beta$ is chosen to assure the correct limit behavior for ballistic motion,
$\beta_{R_0,R_2}(R_1) = p^\ballistic_{R_0,R_2}(R_0|R_1,\leftrightarrow)/p^\ballistic_{R_0,R_2}(R_0|R_1,\leftarrow)$ with $1/4\le\beta\le 1/2$.

For sake of comparison, we note that the result derived in \cite{Vuijk2018} using Fox' colored-noise approximation
can be rewritten in our notation as [see Eq.~(6) in \textit{loc. cit.}]
\begin{equation}
\label{eq:vuijk}
\pi_{R_0,R_2}(R_0|R_1,\leftrightarrow) = 
\frac
{ \displaystyle
  \int_{R_1}^{R_2} \! dR\, \,
	\frac{ p^\ballistic(R_0|R) }{ \alpha l_p(R) } 
}
{ \displaystyle
  \int_{R_0}^{R_2} \! dR\, \,
	\frac{ p^\ballistic(R_0|R) }{ \alpha l_p(R) }  
}.
\end{equation}
Note that for Eq.~(\ref{eq:vuijk}), 
we have mapped the case of klinokinesis with
$\Drot=\Drot(R)$, $v=v_0$ considered in the main text
to the case of orthokinesis with
$\Drot=1/(2\tau_a)$, $v=v(R)$
considered in \cite{Vuijk2018}
using
$v=v_0 / [2\tau_a\,\Drot(R)]$.
This early result is valid for small persistence lengths, 
yet does not provide the correct limit behavior for $R_1\searrow R_0$ and $R_1\nearrow R_2$.
This relates to the fact that the effective Fokker-Planck equation derived in \cite{Vuijk2018} 
is only valid in the limit of small persistence length $l_p\ll R_1-R_0$, $l_p\ll R_2-R_1$.
Note that the authors of \cite{Vuijk2018} had additionally included translational diffusion
with translational diffusion coefficient $D_t$, which we set to zero in Eq.~(\ref{eq:vuijk}) for sake of simplicity.

\begin{figure}
\includegraphics[width=1\linewidth]{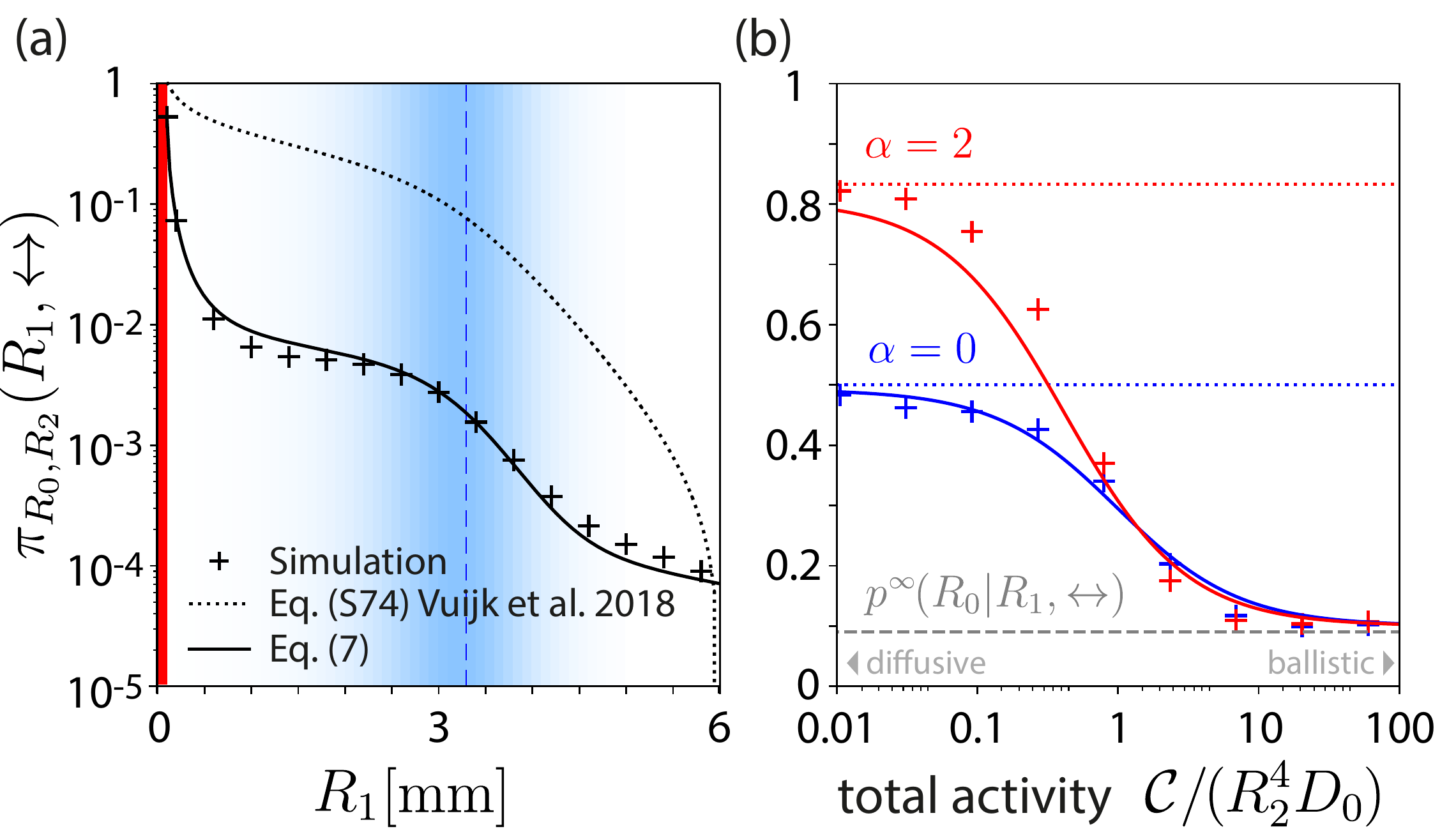}
\caption{
\textbf{Comparison of analytic approximation of target finding probability 
from Vuijk et al. 2018 and Eq.~(7).}
(a)
Splitting probability $\pi_{R_0,R_2}(R_0|R_1,\leftrightarrow)$ of an ABP 
starting at radius $R_1$ with isotropically distributed initial direction
to reach an absorbing boundary at radius $R_0$ instead of an absorbing boundary at radius $R_2$
as function of $R_1$:
simulation (crosses), 
analytical result Eq.~(\ref{eq:vuijk}) based on Fox' colored noise approximation from \cite{Vuijk2018} (dotted), and
analytical result Eq.~(\ref{eq:new_result}) based on Eq.~(7) derived here using matched asymptotics (solid).
We assume a constant speed $v$ and use the position-dependent rotational diffusion coefficient $\Drot(R)$ from Fig.~1.
(b)
Splitting probability $\pi_{R_0,R_2}(R_0|R_1,\leftrightarrow)$ as in panel (a) 
for a position-dependent speed $v(R)\sim \mathcal{C}/R^\alpha$ as in Eq.~(\ref{eq:v_alpha}) and constant $\Drot=D_0$
as function of the activity parameter $\mathcal{C}$ for constant initial position $R_1$:
simulation (crosses), Eq.~(\ref{eq:vuijk}) (dotted), Eq.~(\ref{eq:new_result}) based on Eq.~(7) (solid),
each for two values of the exponent $\alpha$: $\alpha=0$ (blue), $\alpha=2$ (red).
Parameters:
(a) $R_2=6\,\mathrm{mm}$, $\Drot(R)$: see Fig.~1 and SM section on parameters,
(b) $R_0=R_2/5$, $R_1=R_2/3$. 
}
\label{som_fig_vuijk}
\end{figure}

Fig.~\ref{som_fig_vuijk}(a) compares simulation results for
$\pi_{R_0,R_2}(R_0|R_1,\leftrightarrow)$
with the analytical result Eq.~(\ref{eq:vuijk}) from \cite{Vuijk2018},
and the new result Eq.~(7) and its corollary Eq.~(\ref{eq:new_result}),
using the position-dependent profile $\Drot(R)$ of the rotational diffusion coefficient from Fig.~1.
This profile includes regions with low $\Drot$, and hence large persistence length.
Eq.~(\ref{eq:vuijk}) is thus not applicable and its predictions deviate from the simulations.
We can directly assess the dependence of the accuracy of the approximation Eq.~(\ref{eq:vuijk}) 
as function of persistence length:
we consider a position-dependent speed $v(R)$ obeying a power-law with exponent $\alpha$ as in \cite{Vuijk2018}
[see Eq.~(5) in \textit{loc. cit.}]
\begin{equation}
\label{eq:v_alpha}
v(R) = \frac{ \mathcal{C} }{R^\alpha} \left[ \int_{R_0}^{R_2} dR'\, 4\pi\,R'^2\,R'^{-\alpha} \right]^{-1}.
\end{equation}
Fig.~\ref{som_fig_vuijk}(b) shows simulation results for 
$\pi_{R_0,R_2}(R_0|R_1,\leftrightarrow)$
for fixed $R_1$ as function of total activity $\mathcal{C}$ (mean speed times search volume).
Eq.~(\ref{eq:vuijk}) accurately predicts the limit of this probability for low values of the activity $\mathcal{C}$, 
corresponding to the limit of small persistence length.
Eq.~(7) extends this previous result to the entire range of activities, 
spanning the range from small to large persistence lengths.

\subsection{Possible implementation of two-state chemokinesis with biochemical bistable switch}
\label{som_sec_biochemical_implementation}

Chemokinesis with two internal states requires an internal bistable switch.
The actuation of such a bistable switch could be realized by a positive feedback of chemosensation on itself.
Specifically, in biological cells, post-translational modifications of chemoattractant receptors 
(e.g. phosphorylation, methylation) could enhance their sensitivity by positive feedback once the cell has encountered a sufficiently strong stimulus.

\begin{figure}
\includegraphics[width=1\linewidth]{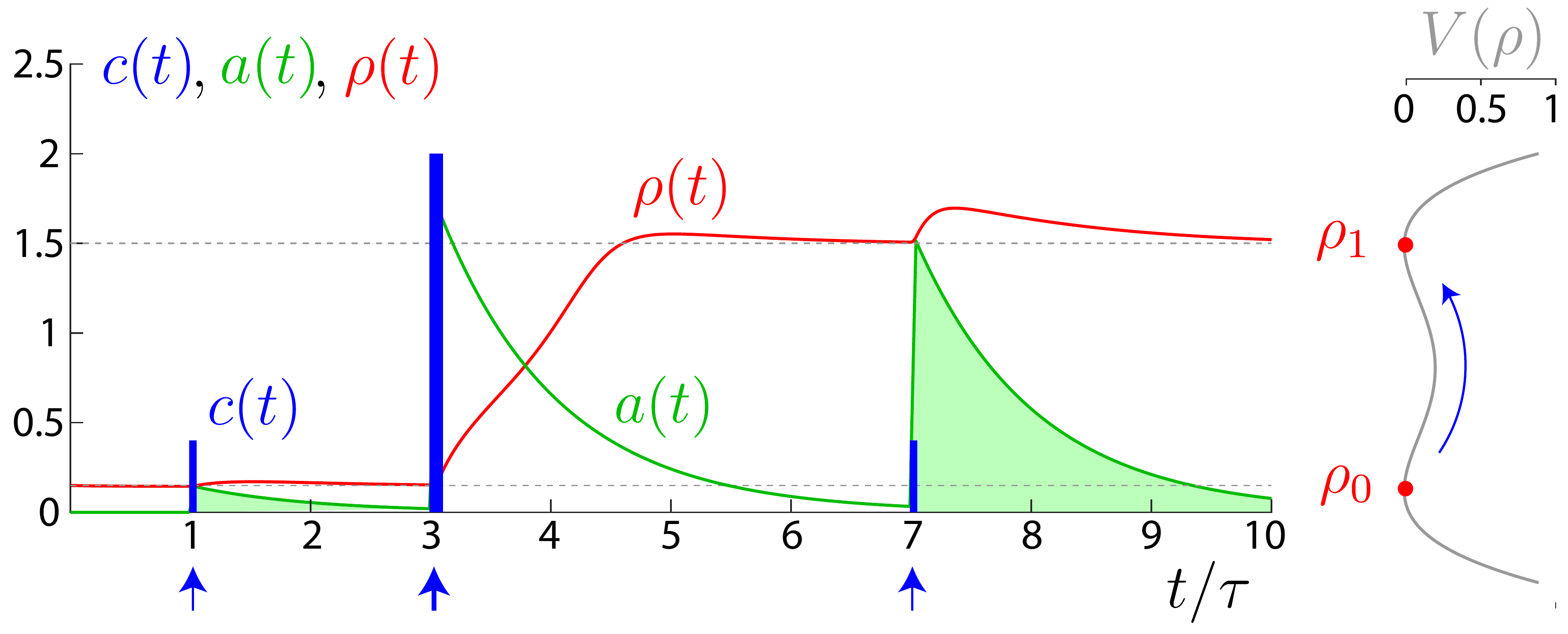}
\caption{
\textbf{Possible implementation of bistable switch for two-state chemokinesis.}
Exemplary numerical solution of the dynamical system Eqs.~(\ref{eq:switch1}), (\ref{eq:switch2}), 
illustrating different responses $a(t)$ (green) to 
identical input pulses $c(t)$ (blue; at times $t/\tau = 1$ and $t/\tau = 7$),
depending on the current value of the dynamic sensitivity $\rho(t)$.
The switch in $\rho(t)$ is triggered by a strong input pulse (blue; at time $t/\tau = 3$).
Parameters: $\rho_1/\rho_0=10$, $\rho_0=0.15$, 
$c(t)=0.2$ for $3\le t/\tau< 3.05$ and $7\le t/\tau< 7.05$, $c(t)=1$ for $6\le t/\tau< 6.1$, $c(t)=0$ else.
}
\label{fig:biochemical_implementation}
\end{figure}

We illustrate this general idea with a minimal model, 
where a cell responds to an extracellular concentration signal $c(t)$ with a time-dependent response $a(t)$ by a simple low-pass filter, 
where the sensitivity $\rho=\rho(t)$ is itself a dynamic variable, 
representing a gain factor that relates input and response
\begin{align}
\label{eq:switch1}
\tau \dot{a} &= \rho c - a , \\
\label{eq:switch2}
\tau \dot{\rho} &= - \frac{d}{d\rho} V(\rho) + a,
\text{ with } V(\rho) = (\rho-\rho_0)^2(\rho-\rho_1)^2 . 
\end{align}
Here, $V(\rho)$ represents a bistable effective potential with two minima for the dynamic sensitivity $\rho$, $\rho_0<\rho_1$.
If the system is initially in state $0$ with $\rho(t=0)=\rho_0$,
it will respond weakly to variations in the input signal $c(t)$. 
A sufficiently strong stimulus, however, can drive the system permanently to state $1$ with $\rho(t) = \rho_1$, 
see Fig.~\ref{fig:biochemical_implementation}.

This minimal model could be generalized to concentration signals 
$c(t)=c(\R(t))$ traced by a chemokinetic agent along its trajectory $\R(t)$
from a spatial concentration field $c(\x)$.
The response variable $a(t)$ could regulate the rotational diffusion coefficient of the agent as
$\Drot = \Drot(a)$.

\subsection{Parameters for Figure 1}
\label{som_sec_parameters}

The concentration field $c(\x)$ and the spatial profile of the signal-to-noise ratio $\mathrm{SNR}$ shown in introductory Fig. 1(a,b) are taken from \cite{Kromer2018}.
The effective rotational diffusion coefficient $\Drot$ shown in Fig.~1(c) is computed according to Eq.~(7) from \textit{loc. cit.}.
For sake of completeness, we include here the details of the computation of 
concentration field $c(\x)$, 
signal-to-noise ratio $\mathrm{SNR}$, and 
rotational diffusion coefficient $\Drot$.

To compute the chemoattractant concentration field $c(\x)$ for the example case of sperm chemotaxis,
we used typical parameters for the sea urchin \textit{Arbacia punctulata}, following \cite{Kromer2018}.
We assume continuous release from a point source for a time $t_c$; hence
\begin{equation}
c = J\, (4 D_c \pi R)^{-1}\, \mathrm{Erfc} \left( \frac{R}{\sqrt{4 D_c t_c}} \right)  .
\end{equation}
Here,
$D_c \approx 239\,\micron^2/s$ is the diffusion coefficient of the chemoattractant resact in sea water \cite{Kashikar2012}, 
and 
$J = n / t_c $ is the release rate of resact (with units of molecules per second), 
where
$n \approx 1.65\cdot 10^{10}$ is the content of resact molecules of a single egg \cite{Kashikar2012}, and
$t_c = 1\,\mathrm{h}$ a typical release time.

The signal-to-noise ratio $\mathrm{SNR}$ of gradient-sensing is computed according to Eq.~(5) of \cite{Kromer2018}
\begin{equation}
\mathrm{SNR} = \frac{(\lambda |\nabla c| r_0)^2 / 2}{\lambda c / T}.
\end{equation}
In fact, this equation is generic and applies up to a prefactor
also to cells performing chemotaxis by spatial comparison \cite{Alvarez2014}.
Here, $r_0$ and $T$ denote a characteristic length-scale and time-scale of gradient sensing, respectively,
and $\lambda$ the binding constant of chemoattractant molecules to receptors on the surface of the cell.
For the specific case of sperm chemotaxis along helical paths, 
we chose 
$r_0 = 7.5\,\micron$ equal to the radius of helical swimming paths, and
$T=0.34\,\mathrm{s}$ equal to the period of helical swimming \cite{Jikeli2015}.
We use the estimate
$\lambda = 7\, \mathrm{pM}^{-1}\,\mathrm{s}^{-1}$
for the binding constant of resact molecules to guanylate cyclase receptors on the surface of the sperm cell \cite{Pichlo2014}. 

We quote Eq.~(7) from \cite{Kromer2018} for the effective rotational diffusion coefficient 
resulting from chemotaxis in the presence of sensing noise
\begin{equation}
\Drot = \left(\frac{\rho\varepsilon}{T}\right)^2 \frac{c}{\lambda (c_b + c)^2 } .
\end{equation}
This equation was derived by coarse-graining the stochastic equations of motion in a model of sperm chemotaxis along helical paths \cite{Friedrich2009}.
Yet, this equation is general and applies to any chemotaxis scenario, 
where a chemotactic agent gradually aligns its net swimming direction to the estimated direction of an external concentration gradient $\nabla c$. 
Here, it is assumed that the chemotactic agent performs sensory adaption with sensitivity threshold $c_b$, 
i.e., the amplitude of chemotactic steering responses scales as
$\rho\, c / (c_b + c)$. 
Here, 
$\rho$ denotes a gain factor that characterizes the amplitude of chemotactic steering responses, and
$\varepsilon$ is a dimensionless geometric factor specific to the case of helical chemotaxis.
We compute the geometric factor $\varepsilon$ according to \cite{Kromer2018} as
$\varepsilon=2\pi\kappa_0\tau_0/(\kappa_0^2+\tau_0^2)$,
using measured values of mean path curvature and mean path torsion of helical sperm swimming paths,
$\kappa_0 = 0.065\,\micron^{-1}$ and 
$\tau_0 = 0.067\,\micron^{-1}$, respectively \cite{Jikeli2015}.
We use
$v=150\,\micron\,\mathrm{s}^{-1}$ for the net swimming speed along the centerline of helical swimming paths,
which corresponds to a speed of $v_0 \approx 200\,\micron\,\mathrm{s}^{-1}$ along the helical path itself, 
according to 
$v = v_0\, \tau_0 / [\kappa_0^2+\tau_0^2]^{1/2}$, 
consistent with previous measurements \cite{Jikeli2015}.
The gain factor $\rho$ is chosen as $\rho=1$.
This value reproduced typical bending rates of helical swimming paths as observed in experiments \cite{Jikeli2015}.
The adaptation threshold is set as
$c_b=10\,\mathrm{pM}$.
At this concentration $c_b$, about $20$ chemoattractant molecules would diffuse to a sperm cell during one helical turn.
Note that sea urchin sperm cells respond to single chemoattractant molecules \cite{Pichlo2014};
the change in intracellular calcium concentration caused by the binding of chemoattractant molecules 
as function of stimulus strength becomes sublinear already for chemoattractant concentrations on the order of $c_b$ \cite{Kashikar2012}.


\end{document}